\documentclass[preprint,12pt]{elsarticle}
\usepackage{graphicx}
\usepackage{amssymb}
\usepackage{amsmath}
\usepackage{subfig}
\usepackage{multirow}
\usepackage{colortbl,color}
\usepackage{algorithm}
\usepackage{amsthm}
\usepackage{hyperref}
\biboptions{comma,sort&compress}
\journal{ }

\begin{document}

\begin{frontmatter}
\title{Comparison between the DSMC and DSBGK Methods}
\author[kaust]{Jun~Li\footnote[1]{e-mail: lijun04@gmail.com \textit{or} jun.li@kaust.edu.sa\\ To view the network videos of DSBGK transient evolution of several benchmark problems, please \href{https://www.dropbox.com/sh/w6t5s7cyteo60fs/Nb5umVXBIU}{\color {red} {click here}} or find the link at \url{http://www.mendeley.com/profiles/junli-cv/} \\ Sincere thanks to my wife Ailing Wang for her constantly supports.}}
\address[kaust]{Applied Mathematics and Computational Science\\
King Abdullah University of Science and Technology\\Thuwal, Saudi Arabia}
\begin{abstract}
Recently, the DSBGK method (\textit{note}: the original name DS-BGK is changed to DSBGK for simplicity) was proposed based on the BGK equation to reduce the stochastic noise in simulating rarefied gas flows at low velocity, in which the deviation from equilibrium state is small making the traditional DSMC simulation time-consuming due to the dominance of noise in transient results. In both DSMC and DSBGK simulations, the simulated molecules move into and out of cells randomly and frequently. Consequently, the transient information of molecules in each particular cell contains significant noise. The DSMC method uses the transient values of molecular variables to compute the cell's variables (including number density, flow velocity and temperature) and so the stochastic noise in its cell's variables is remarkable particularly in the case of low velocity. In the DSBGK simulation, the increments rather than the transient information of molecular variables are used to update the cell's variables based on the mass, momentum and energy conservation principles of intermolecular collision process. This updating scheme significantly reduces the noise in cell's variables of DSBGK simulations because the molecular variables are updated smoothly by the extrapolation of acceptance-rejection scheme and so their increments contain low noise. The detailed comparisons of algorithms and results between the DSMC and DSBGK methods are given here. Several benchmark problems are simulated to verify the DSBGK method by comparison with the DSMC method as criterion.
\end{abstract}
\end{frontmatter}
\section{Introduction}\label{s:intro}
For micro gas flows, the Boltzmann equation rather than the Navier-Stokes equation should be used due to high Knudsen number $Kn=\lambda/L$ where $\lambda$ is the molecular mean free path and $L$ is the characteristic length of the flow problem. In addition, the influence of boundary condition to the solutions becomes dominant because the frequency of molecular reflection on the solid wall, compared to the frequency of intermolecular collision, increases with $Kn$. Unfortunately, the characteristic velocity of micro gas flows is usually much smaller than the molecular random thermal velocity and sometimes the variations of quantities of interest inside the flow domain are very small, which makes the traditional DSMC method \cite{Graeme1994} time-consuming although it is successful in the case of high velocity.

The DSBGK method \cite{Jun2011RGD} was proposed to improve the efficiency in simulating micro gas flows and verified in the lid-driven, Couette and channel flow problems \cite{Jun2011RGD}-\cite{Jun2011ICNMM} by comparison with the DSMC method as criteria. Theoretically, it can be proved, as will be discussed later, that the solution of the DSBGK method converges to the steady-state solution of the BGK equation \cite{BGK1954}. The application of the CLL reflection model \cite{Carlo1971CLL}-\cite{Lord1991CLL} in the DSBGK method is possible and few tentative results were compared with the DSMC results in \cite{Jun2011ICNMM}. Although based on the BGK equation obtained by using a simple model to replace the intermolecular collision integral of the Boltzmann equation, the DSBGK method agrees well with the DSMC method at $Kn=0.063$ and $6.3$ in the lid-driven problem \cite{Jun2011RGD}. This is because the molecular reflection on wall, the dominant effect in micro gas flows, is modeled by the DSBGK method in the same way as by the DSMC method. Theoretically, the error due to simplification to the intermolecular collision process vanishes and the solution depends only on the boundary condition when $Kn\to\infty$. The DSBGK method achieves high efficiency by avoiding generating random fractions in the intermolecular collision process and using the increments (instead of transient values) of molecular variables to update cell's macro quantities, which significantly reduces the statistical noise due to discontinuous events of simulated molecules moving into and out of cells. Consequently, the total computational time used by the DSBGK simulation almost not increase with the decrease of magnitude of the deviation from equilibrium state and sometimes the average process can be avoided as the transient cell's variables contain few stochastic errors \cite{Jun2011RGD}. In addition to its high-efficiency, the DSBGK method has many numerical advantages including simplicity, stability, convenience for complex configuration and for parallel computation because the basic algorithmic structure of the DSMC method is employed.

The comparison between the DSMC and DSBGK algorithms is given here. Theoretical analysis is provided to show the convergence of the DSBGK method to the BGK equation. Then, the results of several benchmark problems, including the Couette flow, channel flow, lid-driven flow and thermal transpiration problem, are listed together to show the agreement of the DSBGK method with the DSMC method. The benchmark problems are divided into closed and open problems to discuss the efficiency and stability of the DSBGK simulation separately. In closed problems, the long-period fluctuation is observed in the number density distribution of DSBGK simulations. Many simulated molecules are employed to reduce the magnitude of fluctuation and improve the numerical stability. Consequently, the memory usage is increased remarkably but the efficiency is still very high as shown in the closed lid-driven problem \cite{Jun2011RGD}. In open problems, the boundary condition with fixed number density eliminates the unphysical fluctuation and the DSBGK simulation remains stable even when using about $10$ simulated molecules per cell, which significantly reduces the memory usage and so improves the applicability in open problems of large scale.
\section{DSMC Method}\label{s:DSMC method}
The DSMC method \cite{Graeme1994}, which is successful in simulating rarefied gas flows at high velocity, was proposed based on physical understanding with appropriate theoretical analysis. In fact, the DSMC algorithm in simple cases can be understood by using the importance sampling scheme to solve the Boltzmann equation \cite{Lowell2005ImportanceSampling}-\cite{Jun2009thesis}, which is discussed here. The rarefied gas flow is described by the Boltzmann equation. We consider gas flows of single component in the absence of external body force. If the molecule is modeled by a hard sphere with fixed diameter $D$, the Boltzmann equation is:
\begin{equation}\label{eq:Boltzmann}
\begin{aligned}
    \dfrac{\partial f}{\partial t}+c_j\dfrac{\partial f}{\partial x_j}&=\dfrac{\partial f}{\partial t}|_{\mathrm{coll}} \\
    &=\dfrac{1}{2}\int_{-\infty}^{\infty}\int_{-\infty}^{\infty}\int_{0}^{4\pi}(\delta_2'+\delta_1'-\delta_2-\delta_1)g\dfrac{D^2}{4}f_1f_2
    \mathrm{d}\Omega \mathrm{d}\vec c_1\mathrm{d}\vec c_2
\end{aligned}
\end{equation}
$f(t, \vec x, \vec c)$ is the unknown probability distribution function, $t$ is the time, $\vec x$ is the spatial coordinate and $\vec c$ is the molecular velocity, $f_1=f(t, \vec x, \vec c_1)$ and $f_2=f(t, \vec x, \vec c_2)$, the delta function $\delta_1=\delta(\vec c_1-\vec c)$, $\delta_2=\delta(\vec c_2-\vec c)$, $\delta_1'=\delta(\vec c_1'-\vec c)$, $\delta_2'=\delta(\vec c_2'-\vec c)$, $g=|\vec c_2-\vec c_1|$, the post-collision velocities $\vec c_1', \vec c_2'$ are determined by the pre-collision velocities $\vec c_1, \vec c_2$ and the solid angle $\Omega$, $\mathrm{d}\Omega=\sin\varphi\mathrm{d}\varphi\mathrm{d}\theta$ where $\varphi\in[0, \pi]$ is the polar angle (the deflection angle in intermolecular collisions) and $\theta\in[0, 2\pi]$ is the azimuthal angle of the spherical coordinate system. The total collision section is $\sigma_\mathrm{T}=\int_{0}^{4\pi}D^2/4\mathrm{d}\Omega=\pi D^2$. The boundary condition will be discussed later in section \ref{ss:boudary condition} together with the DSBGK method. After getting the solution of $f(t,\vec x, \vec c)$, the number density $n(t, \vec x)$, flow velocity $\vec u(t, \vec x)$ and temperature $T(t, \vec x)$ are computed
\begin{equation}\label{eq:nuT}
    \begin{cases}
    n=\int_{-\infty}^{\infty}f\mathrm{d}\vec c \\
    \vec u=\dfrac{\int_{-\infty}^{\infty}\vec cf\mathrm{d}\vec c}{n} \\
    T=\dfrac{\int_{-\infty}^{\infty}\dfrac{m(\vec c-\vec u)^2}{2}f\mathrm{d}\vec c}{\dfrac{3k_\mathrm{B}}{2}n} \\
    \end{cases}
\end{equation}
where $m$ is the molecular mass and $k_\mathrm{B}$ is the Boltzmann constant. Higher order momentums, like shear stress tensor and heat flux, are computed similarly.

In the DSMC simulation \cite{Graeme1994}, each simulated molecule $l$ carries two molecular variables: position $\vec x_l$ and velocity $\vec c_l$. In order to reduce the memory usage, the number of simulated molecules is much smaller than that of the real molecules contained in the flow domain and so we assume that each simulated molecule represents $N$ number of real molecules. Note that $N$ is a constant and very large to make each cell usually containing about 20 simulated molecules. The molecular position and velocity are selected at the initial state and updated during the simulation process appropriately such that the set of all simulated molecules $[\vec x_l, \vec c_l]_{\mathrm{all}}$ represents the probability distribution function $f$ and its evolution with time, which means that the simulated molecules are distributed according to $f$ in the phase space $(\vec x, \vec c)$ at any moment $t$. The flow domain is divided into many cells and $n, \vec u, T$ are estimated by summation inside each cell $k$ using $N/V_k$ to replace $f\mathrm{d}\vec c$ in Eq. (\ref{eq:nuT}) as $f\mathrm{d}\vec c\mathrm{d}\vec x$ is the number of real molecules in the velocity space element $\mathrm{d}\vec c$ and the physical space element $\mathrm{d}\vec x$:
\begin{equation}\label{eq:nuT-discrete}
    \begin{cases}
    n_{k}=\dfrac{\sum N}{V_k} \\
    \vec u_{k}=\dfrac{\sum(N\vec c_l)}{\sum N} \\
    T_{k}=\dfrac{\sum [N\dfrac{m(\vec c_l-\vec u_{k})^2}{2}]}{\dfrac{3k_\mathrm{B}}{2}\sum N} \\
    \end{cases}
\end{equation}
where $V_k$ is the volume of cell $k$, $\sum$ is the summation over those simulated molecules located inside cell $k$ at any particular moment $t$. For example, $\sum N$ is the product of $N$ and the number of simulated molecules and so equal to the number of real molecules inside cell $k$.

During each time step $\Delta t$, we split $\dfrac{\partial f}{\partial t}$ into $\dfrac{\partial f}{\partial t}|_{\mathrm{move}}=-c_j\dfrac{\partial f}{\partial x_j}$ due to free molecular motions and $\dfrac{\partial f}{\partial t}|_{\mathrm{coll}}$ due to intermolecular collisions. As $[\vec x_l, \vec c_l]_{\mathrm{all}}$ is a representative sample of $f$, $\dfrac{\partial f}{\partial t}|_{\mathrm{move}}$ is represented by updating $\vec x_l$ when simulated molecules move uniformly and in a straight line.

For $\dfrac{\partial f}{\partial t}|_{\mathrm{coll}}$, we need to calculate the increment $\Delta f|_{\mathrm{coll}}$ of $f$ after each $\Delta t$ at all spatial points $\vec x$ and all velocity points $\vec c$ inside the whole phase space. In order to make $\Delta f|_{\mathrm{coll}}$ tractable, we assume that the coordinates $\vec x_l$ of those simulated molecules inside the same cell $k$ are the same (notated by $\vec x_{\mathrm{center,}k}$). Then, we only need to compute $\Delta f|_{\mathrm{coll}}$ at those discrete spatial points $\vec x_{\mathrm{center,}k}$ of each cell (as $f=0$ and so $\Delta f=0$ at other spatial points) but still at all velocity points. The distribution function at $\vec x_{\mathrm{center,}k}$ is $f_k=\sum \delta(\vec c_l-\vec c)N/V_k$ which is consistent with Eqs. \eqref{eq:nuT}-\eqref{eq:nuT-discrete} as $n_k=\int_{-\infty}^{\infty}f_k\mathrm{d}\vec c=\sum N/V_k$ (again, $\sum$ is over simulated molecules inside cell $k$). At the end of each $\Delta t$ and for each cell $k$, we compute $\Delta f_k|_{\mathrm{coll}}$ according to the Boltzmann equation:
\begin{equation}\label{eq:Df}
\begin{aligned}
    \Delta f_k|_{\mathrm{coll}}&=\Delta t\dfrac{\partial f_k}{\partial t}|_{\mathrm{coll}} \\
    &=\dfrac{\Delta t}{2}\int_{-\infty}^{\infty}\int_{-\infty}^{\infty}\int_{0}^{4\pi}(\delta_2'+\delta_1'-\delta_2-\delta_1)
    g\dfrac{D^2}{4}f_{k,1}f_{k,2}\mathrm{d}\Omega\mathrm{d}\vec c_1\mathrm{d}\vec c_2 \\
    &=M\int_{-\infty}^{\infty}\int_{-\infty}^{\infty}\int_{0}^{4\pi}
    G\dfrac{D^2}{4\sigma_\mathrm{T}}\mathrm{d}\Omega
    \dfrac{f_{k,1}}{n_k}\mathrm{d}\vec c_1\dfrac{f_{k,2}}{n_k}\mathrm{d}\vec c_2
\end{aligned}
\end{equation}
where $M=\dfrac{\Delta tn_k^2V_k(g\sigma_\mathrm{T})_{\max}}{2N}$ and $G=\dfrac{N}{V_k}(\delta_2'+\delta_1'-\delta_2-\delta_1)\dfrac{g\sigma_\mathrm{T}}{(g\sigma_\mathrm{T})_{\max}}$, $f_{k,1}=\sum \delta(\vec c_l-\vec c_1)N/V_k$ is the distribution function of $\vec c_1$ at $\vec x_{\mathrm{center,}k}$. Note that the value of $g$ has upper bound here as $f_k$ is nonzero only at finitely many discrete velocity points $\vec c_l$. Although the value of $(g\sigma_\mathrm{T})_{\max}$ can be any constant in Eq. \eqref{eq:Df}, it should be updated appropriately by the existing values $g\sigma_\mathrm{T}$ in all cells during each $\Delta t$ in the DSMC simulation such that the ratio $g\sigma_\mathrm{T}/(g\sigma_\mathrm{T})_{\max}$ is always (\textit{note}: practically will be 'almost always') smaller than $1$ which is required by the following acceptance-rejection scheme. But, if $(g\sigma_\mathrm{T})_{\max}$ is much larger than that required to make all ratios smaller than $1$, the number $M$ of tentative collision pairs is very large making the simulation process time-consuming due to low acceptance probabilities of the tentative collisions (see the following analysis). Note that $\int_{0}^{4\pi}\dfrac{D^2}{4\sigma_\mathrm{T}}\mathrm{d}\Omega=1$, $\int_{-\infty}^{\infty}\dfrac{f_{k,1}}{n_k}\mathrm{d}\vec c_1=1$, $\int_{-\infty}^{\infty}\dfrac{f_{k,2}}{n_k}\mathrm{d}\vec c_2=1$ and so, $\Delta f_k|_{\mathrm{coll}}$ is equal to $M<G>$ where $<G>$ is the expected value of $G$. We use the importance sampling scheme to estimate $<G>$, namely $\dfrac{1}{n_{\mathrm{sample}}}\sum_{\mathrm{sample}} G_j\approx <G>$ where $\sum_{\mathrm{sample}} G_j$ is the sum of $n_{\mathrm{sample}}$ number of representative $G_j=G_j(\Omega, \vec c_1, \vec c_2)$ with $\Omega, \vec c_1, \vec c_2$ being selected according to their probability densities $\dfrac{D^2}{4\sigma_\mathrm{T}}, \dfrac{f_{k,1}}{n_k}, \dfrac{f_{k,2}}{n_k}$, respectively. Furthermore, we let $n_{\mathrm{sample}}=M$ and so $\Delta f_k|_{\mathrm{coll}}\approx\sum_{\mathrm{sample}} G_j$.

For any $G_j$, we select particle $j_1$ randomly and uniformly from those simulated molecules inside cell $k$ and thus $\vec c_1=\vec c_{j_1}$ is selected according to $f_{k,1}/n_k$ as required because $f_{k,1}=\sum \delta(\vec c_l-\vec c_1)N/V_k$, which implies that all simulated molecules should be selected equivalently. The number of simulated molecules inside $\mathrm{d}\vec c$ represents $f_k$. Then, we select particle $j_2$ ($j_2\ne j_1$) randomly and uniformly inside cell $k$ and use $\vec c_{j_2}$ as the $j^{th}$ representative value of $\vec c_2$, which also implies that $\vec c_2$ is selected according to $f_{k,2}/n_k$ where $f_{k,2}=\sum \delta(\vec c_l-\vec c_2)N/V_k$. As $D^2/(4\sigma_\mathrm{T})$ is a constant, we select $\Omega$ randomly and uniformly from the whole surface of unit sphere, which is equivalent to selecting the post-collision $\vec c_{j_1}', \vec c_{j_2}'$ randomly by the hard-sphere collision model as $\Omega$ is used only to calculate $\vec c_{j_1}', \vec c_{j_2}'$. Now, we have $\vec c_{j_1}, \vec c_{j_2}, \vec c_{j_1}', \vec c_{j_2}'$ and $g_j=|\vec c_{j_1}-\vec c_{j_2}|$. Assuming that $(g\sigma_\mathrm{T})_{\max}$ is known for the current $\Delta t$, $G_j$ is equal to $\dfrac{N}{V_k}[\delta(\vec c_{j_2}'-\vec c)+\delta(\vec c_{j_1}'-\vec c)-\delta(\vec c_{j_2}-\vec c)-\delta(\vec c_{j_1}-\vec c)]\dfrac{g_j\sigma_\mathrm{T}}{(g\sigma_\mathrm{T})_{\max}}$. Now, the acceptance-rejection scheme is used to handle the fraction $\dfrac{g_j\sigma_\mathrm{T}}{(g\sigma_\mathrm{T})_{\max}}$. If $\dfrac{g_j\sigma_\mathrm{T}}{(g\sigma_\mathrm{T})_{\max}}>Rf$ where $Rf$ is a random fraction distributed uniformly inside [0, 1], we let $G_j=\dfrac{N}{V_k}[\delta(\vec c_{j_2}'-\vec c)+\delta(\vec c_{j_1}'-\vec c)-\delta(\vec c_{j_2}-\vec c)-\delta(\vec c_{j_1}-\vec c)]$ and $G_j=0$ otherwise. Note that $f_k=\sum \delta(\vec c_l-\vec c)N/V_k$ and $\Delta f_k|_{\mathrm{coll}}\approx\sum_{\mathrm{sample}} G_j$ and so $f_k$ becomes $\sum \delta(\vec c_l-\vec c)N/V_k+\sum_{\mathrm{sample}} G_j$ after intermolecular collisions. This implies that if $\dfrac{g_j\sigma_\mathrm{T}}{(g\sigma_\mathrm{T})_{\max}}>Rf$, $\dfrac{N}{V_k}[\delta(\vec c_{j_2}-\vec c)+\delta(\vec c_{j_1}-\vec c)]$ contained in $\sum\dfrac{N}{V_k}\delta(\vec c_l-\vec c)$ is canceled by $\dfrac{N}{V_k}[-\delta(\vec c_{j_2}-\vec c)-\delta(\vec c_{j_1}-\vec c)]$ contained in $G_j$ and meanwhile $\dfrac{N}{V_k}[\delta(\vec c_{j_2}'-\vec c)+\delta(\vec c_{j_1}'-\vec c)]$ contained in $G_j$ is added to $\sum\dfrac{N}{V_k}\delta(\vec c_l-\vec c)$, namely replacing $\dfrac{N}{V_k}[\delta(\vec c_{j_2}-\vec c)+\delta(\vec c_{j_1}-\vec c)]$ by $\dfrac{N}{V_k}[\delta(\vec c_{j_2}'-\vec c)+\delta(\vec c_{j_1}'-\vec c)]$ in $\sum\dfrac{N}{V_k}\delta(\vec c_l-\vec c)$. Till now, the replacement may contribute nothing if we are discussing $\Delta f_k|_{\mathrm{coll}}$ at velocity points $\vec c$ different from $\vec c_{j_1}, \vec c_{j_2}, \vec c_{j_1}', \vec c_{j_2}'$ because both the original $\dfrac{N}{V_k}[\delta(\vec c_{j_2}-\vec c)+\delta(\vec c_{j_1}-\vec c)]$ and the new $\dfrac{N}{V_k}[\delta(\vec c_{j_2}'-\vec c)+\delta(\vec c_{j_1}'-\vec c)]$ are equal to zero at those $\vec c$. So, we consider $\Delta f_k|_{\mathrm{coll}}$ at all velocity points $\vec c$ together and specify that the same set of samples $G_j$ is used to compute $\Delta f_k|_{\mathrm{coll}}$ at all different $\vec c$. Then, if $\dfrac{g_j\sigma_\mathrm{T}}{(g\sigma_\mathrm{T})_{\max}}>Rf$, the contribution of $G_j$ to $\Delta f_k|_{\mathrm{coll}}$ in the whole velocity space is nonzero only at four velocity points and equivalent to changing the velocities $\vec c_{j_1}, \vec c_{j_2}$ to $\vec c_{j_1}', \vec c_{j_2}'$, respectively, which means that a pairwise intermolecular collision happens. So, we select $M$ number of tentative collision pairs for each cell $k$ at the end of each $\Delta t$ and use $\dfrac{g_j\sigma_\mathrm{T}}{(g\sigma_\mathrm{T})_{\max}}$ of each pair $j_1, j_2$ as the acceptance probability to judge whether a pairwise collision happens. This is the algorithm used in the DSMC method.

For dense fluids, the importance sampling scheme was used in \cite{Aldo1997Enskog} to solve the Enskog equation, which is an extension of the Boltzmann equation by considering the intermolecular repulsive force at short distance but still neglecting the intermolecular cohesive force at long distance. The cohesive force is vital in simulating two-phase flows \cite{He2002LBM}. For problems at low velocity, the intermolecular collision integral of the Boltzmann equation is simplified and evaluated by the importance sampling scheme to improve the efficiency in the LVDSMC method \cite{Thomas2007LVDSMC}, which conserve the mass on average. A scheme was proposed in \cite{Jun2010LVDSMC} to conserve the mass strictly.

\section{DSBGK Method}\label{s:DSBGK method}
We consider gas flows of single component. In the absence of external body force, the BGK equation \cite{BGK1954} can be written as a Lagrangian form:
\begin{equation}\label{eq:BGK}
    \dfrac{\mathrm{d}f}{\mathrm{d}t}=\dfrac{\partial f}{\partial t}+c_j\dfrac{\partial f}{\partial x_j}=\upsilon(f_{\mathrm{eq}}-f)
\end{equation}
where $f(t, \vec x, \vec c)$ is the unknown probability distribution function, $t$ is the time, $\vec x$ is the spatial coordinate and $\vec c$ is the molecular velocity, the parameter $\upsilon$ is selected appropriately to satisfy the coefficient of viscosity $\mu$ or heat conduction $\kappa$ \cite{Vincenti1965}:
\begin{equation}\label{eq:upsilon}
    \begin{cases}
    \mu_{\mathrm{BGK}}=\dfrac{nk_\mathrm{B}T}{\upsilon} \\
    \kappa_{\mathrm{BGK}}=\dfrac{5k_\mathrm{B}}{2m}\dfrac{nk_\mathrm{B}T}{\upsilon}
    \end{cases}
\end{equation}
and the Maxwell distribution function $f_{\mathrm{eq}}$ is:
\begin{equation}\label{eq:feq}
    f_{\mathrm{eq}}(t, \vec x, \vec c)=n(\dfrac{m}{2\pi k_\mathrm{B}T})^{3/2}\exp[\dfrac{-m(\vec c-\vec u)^2}{2k_\mathrm{B}T}]
\end{equation}
where $m$ is the molecular mass and $k_\mathrm{B}$ is the Boltzmann constant, the number density $n$, flow velocity $\vec u$ and temperature $T$ are functions of $t$ and $\vec x$ and defined by Eq. \eqref{eq:nuT} using $f$ .

In the DSBGK method \cite{Jun2011RGD}, the simulation process is divided into a series of time steps $\Delta t$ and the flow domain is divided into many cells. The selections of $\Delta t$ and cell size are the same as in the DSMC method when simulating problems of high $Kn$. Many simulated molecules are employed to represent the distribution function $f$ and its evolution with time. The main idea of this method is to track down the evolution of $f$ along enormous molecular trajectories at \textit{constant} velocities, which are selected randomly when simulated molecules are generated or reflected at the boundaries. Each simulated molecules $l$ carries four molecular variables: position $\vec x_l$, molecular velocity $\vec c_l$,  number $N_l$ of real molecules represented by the simulated molecule $l$, and $F_l$ which is equal to the representative value $f(t, \vec x_l, \vec c_l)$ of $f$ at the moment $t$ and point $(\vec x_l, \vec c_l)$ in the phase space. $[\vec x_l, \vec c_l, N_l]_{\mathrm{all}}$ is a (not unique) representative sample of $f$ and $[F_l]_{\mathrm{all}}$ is the representative value of $f$. The compatibility condition, namely $[\vec x_l, \vec c_l, N_l]_{\mathrm{all}}$ and $[F_l]_{\mathrm{all}}$ are related to the same $f$, is required during the simulation process. Note that the evolution of $f$ is due to three factors: free molecular motion, intermolecular collision and molecular reflection on the wall.

For the evolution of $f$ due to free molecular motions and intermolecular collisions, $[F_l]_{\mathrm{all}}$ is changed and then $[\vec x_l, \vec c_l, N_l]_{\mathrm{all}}$ is updated correspondingly by changing $\vec x_l$ and $N_l$ rather than $\vec c_l$. Note that $N_l$ is a constant and $\vec x_l$ is changed alone to represent the evolution of $f$ due to free molecular motions and then $\vec c_l$ is changed randomly to represent the evolution of $f$ due to intermolecular collisions in the DSMC simulation. The position $\vec x_l$ is updated along the trajectory of molecular free motion. $F_l$ is updated with $\vec x_l$ by the Lagrangian description of the BGK equation where $f_{\mathrm{eq}}$ is replaced inside each cell $k$ by the transitional $f_{\mathrm{eq,tr,}k}=n_{\mathrm{tr,}k}(\dfrac{m}{2\pi k_\mathrm{B}T_{\mathrm{tr,}k}})^{3/2}\exp[\dfrac{-m(\vec c-\vec u_{\mathrm{tr,}k})^2}{2k_\mathrm{B}T_{\mathrm{tr,}k}}]$. The cell's variables $n_{\mathrm{tr,}k}, \vec u_{\mathrm{tr,}k}, T_{\mathrm{tr,}k}$ are updated by  $\vec x_l, \vec c_l$ and the increment (instead of transient value) of $N_l$ based on the mass, momentum and energy conservation principles of intermolecular collision process. Note that we use the subscript $\mathrm{tr}$ to distinguish the transitional cell's variables $n_{\mathrm{tr,}k}, \vec u_{\mathrm{tr,}k}, T_{\mathrm{tr,}k}$ from $n_{k}, \vec u_{k}, T_{k}$, which are computed by the transient $\vec x_l, \vec c_l, N_l$ as in Eq. \eqref{eq:nuT-discrete} because $[\vec x_l, \vec c_l, N_l]_{\mathrm{all}}$ is a representative sample of $f$. The increment of $N_l$ is due to the intermolecular collision effect and computed by the extrapolation of acceptance-rejection scheme, which avoids the time-consuming process of frequently generating random fractions and employs the changing information of $F_l$ making the compatibility condition satisfied.

For the evolution of $f$ due to molecular reflection at $\vec x_l$ on the wall, $\vec c_l$ is changed to $\vec c_{l\mathrm{,new}}=\vec c_\mathrm{r}+\vec u_{\mathrm{wall}}$ where $\vec c_\mathrm{r}$ is the reflecting velocity selected randomly in the local Cartesian reference system moving at the wall velocity $\vec u_{\mathrm{wall}}$. But, $N_l$ remains unchanged to conserve mass. Then, $F_l$ is updated to $F_{l\mathrm{,new}}=f(t, \vec x_l, \vec c_{l\mathrm{,new}})$, which also satisfies the compatibility condition.
\subsection{Initialization process}\label{ss:initialization}
At the initial state, the cell variables $n_{\mathrm{tr,}k}, \vec u_{\mathrm{tr,}k}, T_{\mathrm{tr,}k}$ are equal to the initial macro quantities which are usually uniform. The initial molecular position $\vec x_l$ and velocity $\vec c_l$ are selected randomly as in the DSMC simulation and then $F_l$ is equal to $f_{\mathrm{eq,tr,}k}(0, \vec x_l, \vec c_l)$. The initial values $N_{l,t=0}$ of $N_l$ for different simulated molecules are usually the same and selected appropriately such that the total number of simulated molecules, which is equal to $N_{\mathrm{total,real}}/N_{l,t=0}$ where $N_{\mathrm{total,real}}$ is the total number of real molecules, takes a acceptable value. The smaller the value of $N_{l,t=0}$ is, the larger the total number of simulated molecules at the initial state will be.
\subsection{Algorithms for molecular motion and intermolecular collision}\label{ss:inside flow domain}
\begin{figure}[H]
  \centering
  \subfloat[Trajectory division]{\includegraphics[width=0.4\textwidth]{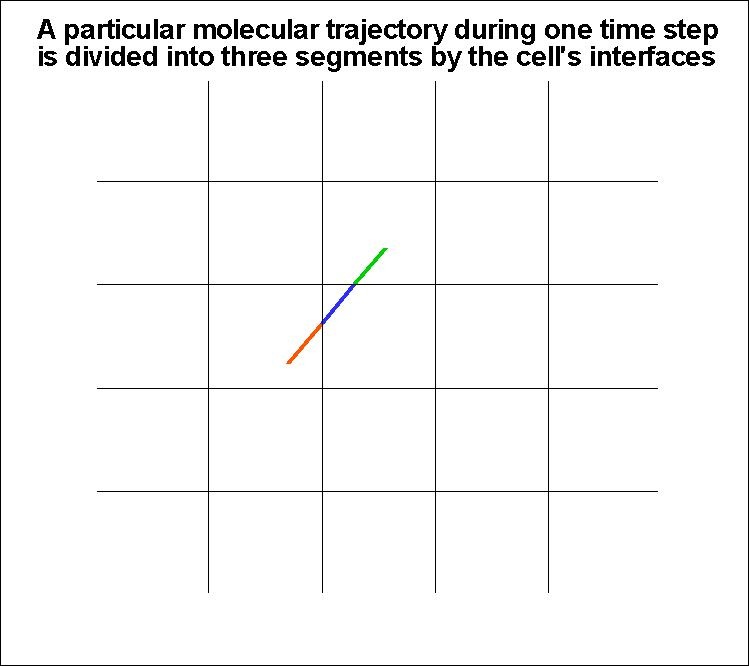}}
  \subfloat[Summation inside each cell]{\includegraphics[width=0.4\textwidth]{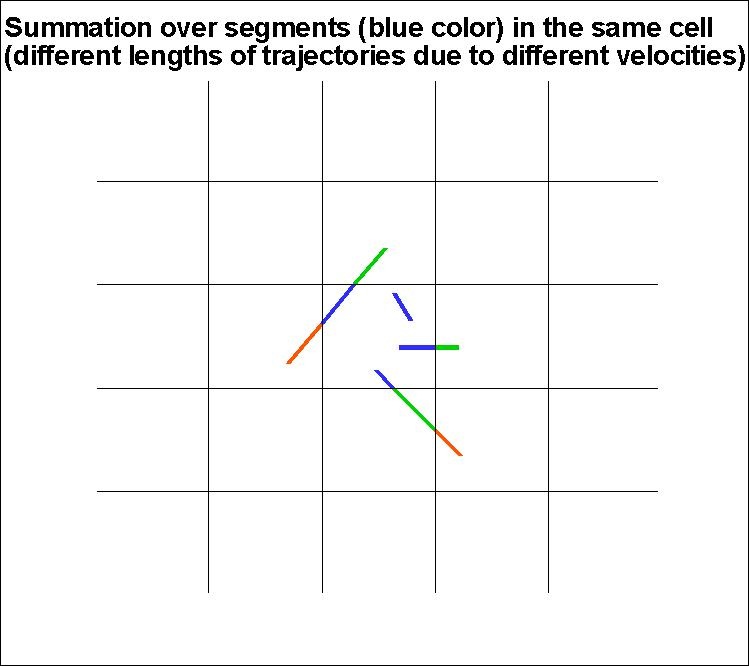}}\\
  \caption{Schematic models of the DSBGK simulation.}
  \label{fig:schematic model}
\end{figure}
In the DSBGK simulation, each simulated molecule moves uniformly and in a straight line before encountering the boundary. As we can see from Fig. \ref{fig:schematic model}, the molecular trajectory during each $\Delta t$ may be divided into several segments by cell's interfaces or remains as a single segment if not yet arriving at any cell's interface at the end of current $\Delta t$. As each segment is located inside a particular cell $k$, $F_l$ is conveniently updated along each segment in sequence according to the Lagrangian form of the BGK equation using $f_{\mathrm{eq,tr,}k}$ of cell $k$. Note that $f_{\mathrm{eq,tr,}k}$ is constant for a particular simulated molecule $l$ and cell $k$ as $n_{\mathrm{tr,}k}, \vec u_{\mathrm{tr,}k}, T_{\mathrm{tr,}k}$ and $\vec c_l$ are fixed. So, $F_l$ is updated using Eq. \eqref{eq:Fnew} obtained by finishing the integration of Eq. \eqref{eq:BGK}, namely $\dfrac{\mathrm{d}F_l}{\mathrm{d}t}=\upsilon(f_{\mathrm{eq,tr,}k}-F_l)$, with respect to $t$ along the segment concerned:
\begin{equation}\label{eq:Fnew}
    F_{l\mathrm{,new}}=f_{\mathrm{eq,tr,}k}+(F_l-f_{\mathrm{eq,tr,}k})\exp(-\upsilon \Delta_kt_l)
\end{equation}
where $F_{l}$ is the previous value and $F_{l\mathrm{,new}}$ is the new value after the intermolecular collision, $\Delta_kt_l$ is the time interval used by the simulated molecule $l$ during the current $\Delta t$ to go through the segment inside cell $k$. As the molecular trajectory is divided first by the time step $\Delta t$ and then by the cell's interfaces, $\Delta_kt_l\leq\Delta t$. If the trajectory during the current $\Delta t$ is divided into several segments by the cell's interfaces, $\Delta_kt_l<\Delta t$ and Eq. \eqref{eq:Fnew} is used repeatedly to update $F_l$ for the consecutive segments in sequence. After updating $F_l$ for each segment, $N_l$ is updated correspondingly:
\begin{equation}\label{eq:Nnew}
    N_{l\mathrm{,new}}=N_lF_{l\mathrm{,new}}/F_l
\end{equation}
which is based on the extrapolation \cite{Jun2011RGD} of acceptance-rejection scheme that if $[\vec x_l, \vec c_l, N_l]_{\mathrm{all}}$ is a representative sample of $f_1$, $[\vec x_l, \vec c_l, N_l(f_2/f_1)_l]_{\mathrm{all}}$ is a representative sample of $f_2$, where $(f_2/f_1)_l$ is the ratio of $f_2$ and $f_1$ at the same point $(t, \vec x_l, \vec c_l)$. Equation \eqref{eq:Nnew} could be understood by considering two steps: in the first step without intermolecular collision, $\vec x_l$ is updated with $t$ along the trajectories but $\vec c_l, F_l, N_l$ keep unchanged as $f(t+\Delta t, \vec x+\Delta t\vec c, \vec c)$=$f(t, \vec x, \vec c)$; then, $F_l$ is changed to $F_{l\mathrm{,new}}$ due to intermolecular collision and $t, \vec x_l, \vec c_l$ keep unchanged, so, $N_l$ is changed correspondingly to $N_{l\mathrm{,new}}$ by Eq. \eqref{eq:Nnew}. The precondition of using the extrapolation of acceptance-rejection scheme is that $[\vec x_l, \vec c_l, N_l]_{\mathrm{all}}$ is a representative sample of $f$ whose representative value is $[F_l]_{\mathrm{all}}$ before intermolecular collision, namely the compatibility condition must holds before using the extrapolation of acceptance-rejection scheme. Then, the updating algorithms of $\vec x_l, F_l, N_l$ with $t$ for the free molecular motion and intermolecular collision processes make the compatibility condition constantly satisfied due to using the extrapolation of acceptance-rejection scheme. In the molecular reflection process on the wall, the compatibility condition is satisfied automatically.

The idea of the updating algorithms along molecular trajectories at \textit{constant} velocities is inspired by the Lattice Boltzmann method (LBM). In turn, the physical understanding of the kinetic equation is also helpful to the development of LBM algorithm. Recently, an alternative scheme was proposed in \cite{Jun2010LBM} to compute the strain rate tensor for the application of large eddy simulation (LES) in the LBM.

The cell's variables $n_{\mathrm{tr,}k}, \vec u_{\mathrm{tr,}k}, T_{\mathrm{tr,}k}$ are used in Eq. \eqref{eq:Fnew} to determine $f_{\mathrm{eq,tr,}k}$ and updated at the end of each $\Delta t$. During the current $\Delta t$ and for each cell $k$ (see Fig. \ref{fig:schematic model} right), some simulated molecules run inside cell $k$ and their increments $\Delta_kN_l=N_{l\mathrm{,new}}-N_l$ inside cell $k$ are already known. $\Delta_kN_l$ is the number increment of real molecules of class $\vec c_l$ associated with the intermolecular collisions inside cell $k$ during the current time step. We make summation $\sum\Delta_kN_l$ over those simulated molecules running inside cell $k$ during the current $\Delta t$ (\textit{note}: simulated molecule $l$ may contribute more than one term to the summation if it reflects on the wall back into the cell $k$). Note that $\Delta_kN_l$ in this summation is the increment information rather than transient information in the summation of Eq. \eqref{eq:nuT-discrete} used in the DSMC method. Obviously, $\sum\Delta_kN_l$ means the number increment of real molecules of all existing classes associated with the intermolecular collisions inside the same cell $k$ during the same time step. So, $\sum\Delta_kN_l$ is expected to be zero as required by the mass conservation principle. Usually, this summation is not exactly equal to zero due to numerical error. So, we decrease $n_{\mathrm{tr,}k}$ if $\sum\Delta_kN_l$ is positive and then $\sum\Delta_kN_l$ will decrease at the next $\Delta t$ as each term $\Delta_kN_l$ decreases due to Eqs. \eqref{eq:Fnew}-\eqref{eq:Nnew}, and vice versa. It works as an auto-regulation scheme which makes $\sum\Delta_kN_l$ approaching to zero. Similarly, $\sum(\Delta_kN_lm\vec c_l)$ and $\sum(\Delta_kN_lm\vec c_l^2/2)$ are related respectively to the momentum increment and kinetic energy increment of real molecules of all existing classes associated with the intermolecular collisions inside the same cell $k$ and during the same $\Delta t$. They are expected to be zero according to the momentum and energy conservation principles of intermolecular collision process and so can be used to update $\vec u_{\mathrm{tr,}k}$ and $T_{\mathrm{tr,}k}$ by auto-regulation schemes. The auto-regulation schemes are:
\begin{equation}\label{eq:auto}
    \begin{cases}
    n_{\mathrm{tr,}k}^{\mathrm{new}}=\dfrac{n_{\mathrm{tr,}k}V_k-\sum\Delta_kN_l}{V_k} \\
    \vec u_{\mathrm{tr,}k}^{\mathrm{new}}=\dfrac{n_{\mathrm{tr,}k}V_k\vec u_{\mathrm{tr,}k}-\sum(\Delta_kN_l\vec c_l)}{n_{\mathrm{tr,}k}^{\mathrm{new}}V_k} \\
    T_{\mathrm{tr,}k}^{\mathrm{new}}=\dfrac{[n_{\mathrm{tr,}k}V_k(\dfrac{3k_\mathrm{B}T_{\mathrm{tr,}k}}{2}+\dfrac{m\vec u_{\mathrm{tr,}k}^2}{2})-\sum(\Delta_kN_l\dfrac{m\vec c_l^2}{2})]-n_{\mathrm{tr,}k}^{\mathrm{new}}V_k\dfrac{m(\vec u_{\mathrm{tr,}k}^{\mathrm{new}})^2}{2}}{n_{\mathrm{tr,}k}^{\mathrm{new}}V_k\dfrac{3k_\mathrm{B}}{2}}
    \end{cases}
\end{equation}
where $n_{\mathrm{tr,}k}^{\mathrm{new}}, \vec u_{\mathrm{tr,}k}^{\mathrm{new}}, T_{\mathrm{tr,}k}^{\mathrm{new}}$ are the new values of number density $n_{\mathrm{tr,}k}$, flow velocity $\vec u_{\mathrm{tr,}k}$ and temperature $T_{\mathrm{tr,}k}$ of cell $k$, respectively, $V_k$ is the volume of cell $k$. The updating schemes of Eq. \eqref{eq:auto} make $\sum\Delta_kN_l$, $\sum(\Delta_kN_l\vec c_l)$, $\sum(\Delta_kN_lm\vec c_l^2/2)$ converging to zero and then $n_{\mathrm{tr,}k}, \vec u_{\mathrm{tr,}k}, T_{\mathrm{tr,}k}$ will fluctuate around their steady state solutions due to stochastic effect.

We use $N_l, F_l$ to represent the previous values at the origin of the segment located inside cell $k$ during the current $\Delta t$ and use $N_{l\mathrm{,new}}, F_{l\mathrm{,new}}$ for the new values at the end of that segment after intermolecular collision as in Eqs. \eqref{eq:Fnew}-\eqref{eq:Nnew}. Note that any possible representative trajectory is selected according to its probability (see section \ref{ss:boudary condition}) as the molecular reflecting velocity is selected randomly according to the boundary reflection model. Thus, it can be expected that the feature of all existing classes represents the feature of all possible classes and so the summation over all existing classes is equivalent to the integration over all possible classes like replacing Eq. \eqref{eq:nuT} by Eq. \eqref{eq:nuT-discrete} in the DSMC simulation. We replace $N_l$ by $V_kF_l\mathrm{d}\vec c_l$ where $\mathrm{d}\vec c_l$ is the velocity space element around $\vec c_l$ as the compatibility condition is satisfied. Note that $\Delta_kt_l$ is the time interval used by the simulated molecule $l$ inside cell $k$ during the current $\Delta t$ and so $\Delta_kt_l=\Delta t$ for those simulated molecules moving inside the same cell (namely the trajectory during the current $\Delta t$ is a single segment without division by the cell's interfaces). We assume that $\Delta t$ is very small making most simulated molecules moving inside the same cell during each $\Delta t$ and so $\Delta_kt_l\simeq\Delta t$. The integral expression of mass conservation of the DSBGK simulation for each cell $k$ is:
\begin{equation}\label{eq:converge of n}
\begin{aligned}
    \sum\Delta_kN_l&=\sum N_{l\mathrm{,new}}-\sum N_l \\
    &=\sum (V_kF_{l\mathrm{,new}}\mathrm{d}\vec c_l)-\sum (V_kF_l\mathrm{d}\vec c_l) \\
    &\approx V_k\sum(\dfrac{\mathrm{d}F_l}{\mathrm{d}t}\Delta_kt_l\mathrm{d}\vec c_l) \\
    &=V_k\sum[\upsilon(f_{\mathrm{eq,tr,}k}-F_l)\Delta_kt_l\mathrm{d}\vec c_l] \\
    &\simeq V_k\Delta t\int_{-\infty}^{\infty}\upsilon(f_{\mathrm{eq,tr,}k}-f)\mathrm{d}\vec c
\end{aligned}
\end{equation}
where $\dfrac{\mathrm{d}F_l}{\mathrm{d}t}=\upsilon(f_{\mathrm{eq,tr,}k}-F_l)$ and the last approximate equality holds as $\Delta_kt_l\simeq\Delta t$ and $F_l$ is the representative value of $f$. So, after convergence with $\sum\Delta_kN_l=0$, we get
\begin{equation}\label{eq:intergral converge of n}
    \int_{-\infty}^{\infty}\upsilon(f_{\mathrm{eq,tr,}k}-f)\mathrm{d}\vec c=0
\end{equation}
But, it is not necessary to require $\Delta t$ being very small. If $\Delta t$ is large, $\sum\Delta_kN_l=0$ still implies $\int_{-\infty}^{\infty}\upsilon(f_{\mathrm{eq,tr,}k}-f)\mathrm{d}\vec c=0$ as both of them represent the mass conservation of intermolecular collision process of the same evolution equation $\dfrac{\mathrm{d}f}{\mathrm{d}t}=\upsilon(f_{\mathrm{eq,tr,}k}-f)$. After convergence with $\sum(\Delta_kN_l\vec c_l)=0$ and $\sum(\Delta_kN_lm\vec c_l^2/2)=0$, the integral expressions of momentum and energy conservations can be obtained similarly. So, the following equation is satisfied for each cell $k$ after convergence:
\begin{equation}\label{eq:converge}
    \int_{-\infty}^{\infty}\upsilon (f_{\mathrm{eq,tr,}k}-f)\psi_i\mathrm{d}\vec c=0
\end{equation}
where $\psi_1=1, (\psi_2, \psi_3, \psi_4)=\vec c, \psi_5=m\vec c^2/2$. As the original BGK equation satisfies $\int_{-\infty}^{\infty}\upsilon (f_{\mathrm{eq}}-f)\psi_i\mathrm{d}\vec c$$=0$, we have $\int_{-\infty}^{\infty}(f_{\mathrm{eq,tr,}k}-f_{\mathrm{eq}})\psi_i\mathrm{d}\vec c=0$ which implies that $n_{\mathrm{tr,}k}=n$, $\vec u_{\mathrm{tr,}k}=\vec u$, $T_{\mathrm{tr,}k}=T$ for each cell $k$ after convergence according to the definitions of $f_{\mathrm{eq,tr,}k}$ and $f_{\mathrm{eq}}$.

So, the solutions of $n_{\mathrm{tr,}k}, \vec u_{\mathrm{tr,}k}, T_{\mathrm{tr,}k}$ of the DSBGK method are the discrete solutions of $n, \vec u, T$ of the BGK equation after convergence under the same boundary condition. Then, the transitional $f_{\mathrm{eq,tr,}k}$ used in the DSBGK method is equal to the original $f_{\mathrm{eq}}$ of the BGK equation inside each cell $k$. Consequently, $[F_l]_{\mathrm{all}}$ and $[\vec x_l, \vec c_l, N_l]_{\mathrm{all}}$ are the representative value and sample, respectively, of the solution $f$ of the BGK equation, which implies that any higher-order moment, including stress tensor and heat flux, calculated by the DSBGK method agrees with that obtained by solving the BGK equation using other numerical methods as in \cite{Zhihui2007Unified} among others.

Note that the updating scheme of Eq. \eqref{eq:auto} conserve the 'total' value $n_{\mathrm{tr,}k}V_k+\sum N_l$ inside each cell $k$ as $(n_{\mathrm{tr,}k}^{\mathrm{new}}V_k-n_{\mathrm{tr,}k}V_k)+\sum\Delta_kN_l=0$ (\textit{note}: 'total' with quotation marks means the sum of cell quantity and molecular quantity). So, $\sum_{\mathrm{Domain}}n_{\mathrm{tr,}k}V_k+\sum_{\mathrm{Domain}}N_l$ is constant during the simulation process because $n_{\mathrm{tr,}k}$ and $N_l$ are unchanged during the molecular reflection process on the wall (\textit{note}: the summation $\sum_{\mathrm{Domain}}$ is over the whole flow domain, namely over all cells and all simulated molecules, respectively). The 'total' momentum and energy of simulated molecules and cells are unchanged when using Eq. \eqref{eq:auto} but not conserved during the whole simulation process due to molecular reflections on the wall, which conserve the mass but not momentum and energy. Note that the conservations of the 'total' mass, momentum and energy by the updating scheme of Eq. \eqref{eq:auto} are artificial restrictions. Eq. \eqref{eq:auto} can be modified by adding arbitrary different \textit{positive} factors before $\sum\Delta_kN_l$, $\sum(\Delta_kN_l\vec c_l)$, $\sum(\Delta_kN_lm\vec c_l^2/2)$ to regulate the convergence speed in open problems. But, the 'total' mass should be conserved in \textit{closed} problems such that
\begin{equation}\label{eq:definite condition}
\begin{aligned}
    \sum_{\mathrm{Domain}}^{\mathrm{Converge}}N_l&{\eqref{eq:intergral converge of n} \atop =}\sum_{\mathrm{Domain}}^{\mathrm{Converge}}n_{\mathrm{tr,}k}V_k
    =\dfrac{1}{2}(\sum_{\mathrm{Domain}}^{\mathrm{Converge}}N_l+\sum_{\mathrm{Domain}}^{\mathrm{Converge}}n_{\mathrm{tr,}k}V_k) \\
    &{\eqref{eq:auto} \atop =}\dfrac{1}{2}(\sum_{\mathrm{Domain}}^{\mathrm{Initial}}N_l+\sum_{\mathrm{Domain}}^{\mathrm{Initial}}n_{\mathrm{tr,}k}V_k)
    =\sum_{\mathrm{Domain}}^{\mathrm{Initial}}N_l=N_{\mathrm{total,real}}
\end{aligned}
\end{equation}
which satisfies the important definite condition for closed problems that the total number $\sum_{\mathrm{Domain}}^{\mathrm{Converge}}N_l$ of real molecules represented by the simulated molecules after convergence is equal to the total number $N_{\mathrm{total,real}}$ of real molecules in the closed physical problem (\textit{note}: total here means the summation $\sum_{\mathrm{Domain}}$ over the flow domain).

Now, we explain why the cell's variables $n_{\mathrm{tr,}k}, \vec u_{\mathrm{tr,}k}, T_{\mathrm{tr,}k}$ are updated by the auto-regulation schemes of Eq. \eqref{eq:auto} rather than Eq. \eqref{eq:nuT-discrete}. As we can see, $F_l$ is updated smoothly by Eq. \eqref{eq:Fnew} and so the increment $\Delta_kN_l$ calculated by Eq. \eqref{eq:Nnew} is also smooth, which implies that the summations $\sum\Delta_kN_l$, $\sum(\Delta_kN_l\vec c_l)$, $\sum(\Delta_kN_lm\vec c_l^2/2)$ used in Eq. \eqref{eq:auto} contain low stochastic noise. But, the summations $\sum N_l$, $\sum(N_l\vec c_l)$, $\sum(N_lm\vec c_l^2/2)$ over transient values as in Eq. \eqref{eq:nuT-discrete} still have large stochastic noise due to the discontinuous events of simulated molecules moving into and out of cell $k$.

The DSBGK algorithm described here is valid for any cell division using parallelepiped or tetrahedron. In the DSMC simulation of problems with complex configuration, we prefer to use the regular parallelepiped to divide the flow domain as in~\cite{Jun2009RGD}, which makes it efficient to determine which cell the simulated molecules are located inside at the end of each $\Delta t$. Although the use of parallelepiped makes it time-consuming to determine the molecular reflection position on the complex wall surface, the number of simulated molecules running into the surface during each $\Delta t$ is usually much smaller than the total number when $Kn$ is much smaller than 1. Compared to using tetrahedrons to divide the flow domain which makes the determination of surface reflection positions of \textit{few} simulated molecules efficient but the determination of the situated cells of \textit{all} simulated molecules after each time step time-consuming, the gain of the algorithm of using parallelepiped outweighs its loss. But, in the DSBGK simulation, the efficiency of the algorithm of molecular motion and intermolecular collision processes depends less on the cell type because the molecular trajectories are divided into segments by cell's interfaces and the molecular variables are updated along each segment in sequence. If molecular reflections on the wall are very frequent and complex wall configurations are involved, we suggest to use tetrahedron to divide the flow domain in the DSBGK simulation such that the determination of surface reflection positions is efficient.

In the DSMC simulation, the total CPU time is almost proportional to the product of sample size $n_{\mathrm{sample}}$ and sampling interval $d_{\mathrm{sample}}$ as the CPU time used for the transitional period before reaching the steady state is usually negligible. The molecular quantities of interest are sampled at intervals ($d_{\mathrm{sample}}=4 \Delta t$ for instance) to reduce the sample size $n_{\mathrm{sample}}$. We use notations $V_1'=V'(\mathrm{CPUtime}, d_{\mathrm{sample,}1})$ and $V_2'=V'(\mathrm{CPUtime}, d_{\mathrm{sample,}2})$ to represent the variances using different $d_{\mathrm{sample}}$ but the same CPU time, namely $d_{\mathrm{sample,}1}\times n_{\mathrm{sample,}1}=d_{\mathrm{sample,}2}\times n_{\mathrm{sample,}2}$. Let $d_{\mathrm{sample,}2}>d_{\mathrm{sample,}1}$ and so $n_{\mathrm{sample,}2}<n_{\mathrm{sample,}1}$. Then, the general rule \cite{Jun2012MCMC} is that $1<\dfrac{V_2'}{ V_1'}\leq\dfrac{n_{\mathrm{sample,}1}}{n_{\mathrm{sample,}2}}$ and the ratio of variance approaches to $1$ when the correlation degree of sample set is very high. So, the increase of statistical variance due to the increase of $d_{\mathrm{sample}}$ from $1$ to $4$ under the conditions of same CPU time is negligible because the correlation degree of consecutive transient results in the DSMC simulation is high. In the DSBGK simulation, the stochastic error is low and the sample size required to obtain smooth results is small. We prefer to sample $n_{\mathrm{tr,}k}, \vec u_{\mathrm{tr,}k}, T_{\mathrm{tr,}k}$ at every time step ($d_{\mathrm{sample}}=1$) as the variance always (although maybe slightly due to high correlation degree) decreases with the increase of sample size under the conditions of same CPU time due to $1<\dfrac{V_2'}{ V_1'}$. Note that $\dfrac{V_2'}{ V_1'}$ approaches to $\dfrac{n_{\mathrm{sample,}1}}{n_{\mathrm{sample,}2}}$ if the consecutive samples are almost independent, which means that the variance is inversely proportional to the sample size and independent of the sampling interval.
\subsection{External body force}\label{ss:external force}
When considering external body force, the BGK equation is changed to:
\begin{equation}\label{eq:BGK-force}
    \dfrac{\partial f}{\partial t}+c_j\dfrac{\partial f}{\partial x_j}+a_j\dfrac{\partial f}{\partial c_j}=\upsilon(f_{\mathrm{eq}}-f)
\end{equation}
where $\vec a=\vec a(t, \vec x)$ is the acceleration due to external body force. We split $\dfrac{\partial f}{\partial t}$ into $\dfrac{\partial f}{\partial t}|_{\mathrm{move}}=-c_j\dfrac{\partial f}{\partial x_j}$, $\dfrac{\partial f}{\partial t}|_{\mathrm{coll}}=\upsilon(f_{\mathrm{eq}}-f)$ and $\dfrac{\partial f}{\partial t}|_{\mathrm{force}}=-a_j\dfrac{\partial f}{\partial c_j}$. To simplify the algorithm, we decouple the effect due to $\dfrac{\partial f}{\partial t}|_{\mathrm{force}}$ from the other two effects. At the end of each $\Delta t$ of the above DSBGK algorithm, the effects due to $\dfrac{\partial f}{\partial t}|_{\mathrm{move}}$ and $\dfrac{\partial f}{\partial t}|_{\mathrm{coll}}$ are already incorporated into the simulation and so we consider $\dfrac{\partial f}{\partial t}|_{\mathrm{force}}$ by changing $\vec c_l$ of each simulated molecule to $\vec c_l+\Delta t\vec a$ and keeping $\vec x_l, F_l, N_l$ unchanged as $f(t+\Delta t, \vec x+\Delta t\vec c, \vec c+\Delta t\vec a)$=$f(t, \vec x, \vec c)$ if neglecting intermolecular collision. Correspondingly, $\vec u_{\mathrm{tr,}k}$ of each cell is changed to $\vec u_{\mathrm{tr,}k}+\Delta t\vec a$ and $n_{\mathrm{tr,}k}, T_{\mathrm{tr,}k}$ keep unchanged. When sampling and outputting the cell's velocity, we use the average value before and after implementing $\vec a$, namely $\vec u_{\mathrm{tr,}k}+0.5\Delta t\vec a$.
\subsection{Boundary conditions}\label{ss:boudary condition}
For the open boundary, simulated molecules are removed from the flow domain when moving across the open boundary during each $\Delta t$. Correspondingly, some new simulated molecules are generated at the end of each $\Delta t$ at the open boundary with $\vec x_l$ and $\vec c_l$ being selected randomly as in DSMC simulations. Then, $F_l$ is determined from $\vec x_l, \vec c_l$ through $f_{\mathrm{eq,tr,}k}$ using the macro quantities fixed at the open boundary or the values of adjacent cell if not prescribed at the boundary. The initial values of $N_l$ of new simulated molecules at different parts of the open boundary can be different in the DSBGK simulation. In the channel flow problem driven by the density difference $\Delta n_{\mathrm{end}}$ at the two ends \cite{Jun2011ICNMM}, we use different initial values of $N_{l\mathrm{,init,end}}$ for different ends such that their ratios of $N_{l\mathrm{,init,end}}/n_{\mathrm{end}}$ are equal, which makes the number of simulated (not real) molecules per cell almost the same for different cells having the same volume but different number density of real molecules. As the stochastic noise at each cell depends on the average number of simulated molecules inside that cell, such selection of the initial values of $N_l$ for new simulated molecules at different parts of the open boundary achieves the trade-off of stochastic noise among cells and so reduces the sample size required for getting smooth results in the whole flow domain.

For the wall boundary, $\vec c_l$ and then $F_l$ are changed after molecular reflection at $\vec x_l$ on the wall as discussed below.
\subsubsection{Updating $\vec c_l$}\label{sss:updating c_l}
When running into the wall and reflecting at $\vec x_l$ on the wall, $\vec c_l$ is changed to $\vec c_{l\mathrm{,new}}=\vec c_\mathrm{r}+\vec u_{\mathrm{wall}}$ where $\vec u_{\mathrm{wall}}$ is the wall velocity and the reflecting velocity $\vec c_\mathrm{r}$ is selected randomly according to the reflection model (namely the scatter kernel discussed later in section \ref{sss:updating F_l}) as in the DSMC simulation. $N_l$ remains unchanged to conserve the mass. After changing $\vec c_l$ alone, $[\vec x_l, \vec c_l, N_l]_{\mathrm{all}}$ is updated to represent $f$ after molecular reflection effect and consequently $F_l$ is updated to the representative value of $f$ at the point $(t, \vec x_l, \vec c_{l\mathrm{,new}})$. So, the compatibility condition is satisfied in the molecular reflection process. The subscript $l$ is omitted in the component expression of velocity when discussing the boundary condition. We predetermine a local Cartesian reference system $S_{\mathrm{local}}$ moving at the wall velocity $\vec u_{\mathrm{wall}}$. We use the subscripts 2 and 3 for the tangential directions and 1 for the normal direction of $S_{\mathrm{local}}$. In the discussion of reflection process, the subscripts 1, 2, 3 always represent the corresponding components in $S_{\mathrm{local}}$. The incoming velocity $\vec c_\mathrm{i}$ is $\vec c_l-\vec u_{\mathrm{wall}}$ (\textit{note}: $\vec c_{l\mathrm{,new}}-\vec u_{\mathrm{wall}}=\vec c_\mathrm{r}$). As $\vec c_l$ and $\vec u_{\mathrm{wall}}$ are stored in the component form of the unique global Cartesian reference system $S_{\mathrm{global}}$, we need the transformation from $S_{\mathrm{global}}$ to $S_{\mathrm{local}}$ to obtain the components of $c_{\mathrm{i,}1}, c_{\mathrm{i,}2}, c_{\mathrm{i,}3}$. Finally, $c_{\mathrm{r,}1}, c_{\mathrm{r,}2}, c_{\mathrm{r,}3}$ are transformed from $S_{\mathrm{local}}$ to $S_{\mathrm{global}}$ to obtain the component form of $\vec c_{l\mathrm{,new}}$ in $S_{\mathrm{global}}$. For the unit normal vector $\vec e_{\mathrm{n}}$ of wall, we specify that the selection of $\vec e_{\mathrm{n}}$ makes the incoming component $c_{\mathrm{i,}1}=\vec c_\mathrm{i}\cdot\vec e_{\mathrm{n}}$ negative and the reflecting component $c_{\mathrm{r,}1}=\vec c_\mathrm{r}\cdot\vec e_{\mathrm{n}}$ positive. The normal direction is unique and the selections of tangential directions are free but fixed during the simulation process. In the original CLL reflection model \cite{Carlo1971CLL}-\cite{Lord1991CLL}, we compute the tangential components of $\vec c_\mathrm{r}$ by $c_{\mathrm{r,}2}=v\cos\theta-w\sin\theta$ and $c_{\mathrm{r,}3}=v\sin\theta+w\cos\theta$ where $v=[(1-\alpha_\tau)(c_{\mathrm{i,}2}^2+c_{\mathrm{i,}3}^2)]^{1/2}+(2k_\mathrm{B}T_{\mathrm{wall}}/m)^{1/2}r_{\tau}\cos\varphi_{\tau}$, $w=(2k_\mathrm{B}T_{\mathrm{wall}}/m)^{1/2}r_{\tau}\sin\varphi_{\tau}$, $r_{\tau}=(-\alpha_{\tau}\ln Rf_1)^{1/2}$, $\varphi_{\tau}=2\pi Rf_2$, $\theta$ is the azimuthal angle of incoming velocity component $(c_{\mathrm{i,}2}, c_{\mathrm{i,}3})$ in the $x_2x_3$ plane of $S_{\mathrm{local}}$, $Rf_1$ and $Rf_2$ are two different random fractions distributed uniformly inside [0, 1], $\alpha_{\tau}$ is the accommodation coefficient of kinetic energy of the tangential velocity component. For the normal component, $c_{\mathrm{r,}1}=[(1-\alpha_\mathrm{n})c_{\mathrm{i,}1}^2+(2k_\mathrm{B}T_{\mathrm{wall}}/m)r_\mathrm{n}^2+2(1-\alpha_\mathrm{n})^{1/2}|c_{\mathrm{i,}1}|
(2k_\mathrm{B}T_{\mathrm{wall}}/m)^{1/2}r_\mathrm{n}\cos\varphi_\mathrm{n}]^{1/2}$ where $|c_{\mathrm{i,}1}|$ is the absolute value of $c_{\mathrm{i,}1}$ as $c_{\mathrm{i,}1}<0$, $r_\mathrm{n}=(-\alpha_\mathrm{n}\ln Rf_3)^{1/2}$, $\varphi_\mathrm{n}=2\pi Rf_4$, $Rf_3, Rf_4$ are two additional random fractions and $\alpha_\mathrm{n}$ is the accommodation coefficient of kinetic energy of the normal velocity component.

We get $c_{\mathrm{r,}2}=c_{\mathrm{i,}2}(1-\alpha_{\tau})^{1/2}+(2k_\mathrm{B}T_{\mathrm{wall}}/m)^{1/2}r_{\tau}\cos(\varphi_{\tau}+\theta)$ and $c_{\mathrm{r,}3}=c_{\mathrm{i,}3}(1-\alpha_{\tau})^{1/2}+(2k_\mathrm{B}T_{\mathrm{wall}}/m)^{1/2}r_{\tau}\sin(\varphi_{\tau}+\theta)$ after reorganizing the formulas of $c_{\mathrm{r,}2}, c_{\mathrm{r,}3}$. Note that $\varphi_{\tau}$ is selected uniformly from a periodic interval $[0, 2\pi]$ and so $\varphi_{\tau}+\theta$ can be replaced simply by $\varphi_{\tau}$, which implies that the calculation of $\theta$ can be avoided to slightly improve the efficiency. So, for the CLL reflection model, the equivalent but simpler algorithm to compute the tangential components in $S_{\mathrm{local}}$ is that $c_{\mathrm{r,}2}=c_{\mathrm{i,}2}(1-\alpha_{\tau})^{1/2}+(2k_\mathrm{B}T_{\mathrm{wall}}/m)^{1/2}r_{\tau}\cos\varphi_{\tau}$ and $c_{\mathrm{r,}3}=c_{\mathrm{i,}3}(1-\alpha_{\tau})^{1/2}+(2k_\mathrm{B}T_{\mathrm{wall}}/m)^{1/2}r_{\tau}\sin\varphi_{\tau}$ \cite{Jun2011ICNMM}. This simpler algorithm also degenerates to the Maxwell diffuse reflection model when $\alpha_{\tau}=\alpha_\mathrm{n}=1$.
\subsubsection{Updating $F_l$}\label{sss:updating F_l}
After getting $\vec c_\mathrm{r}$, $F_l$ is updated correspondingly to $F_{l\mathrm{,new}}=f(t, \vec x_l, \vec c_{l\mathrm{,new}})=f(t, \vec x_l, \vec c_\mathrm{r}+\vec u_{\mathrm{wall}})$. Note that $F_l$ is the representative value of $f$ which is different from the scatter kernel $R$ used to select $\vec c_\mathrm{r}$ for each particular reflection process. Generally speaking, $f$ is related to the mass flux but $R$ has nothing to do with the mass flux. Usually, $R$ describes the distribution probability of $\vec c_\mathrm{r}$ inside the half velocity space ($\vec c_\mathrm{r}\cdot\vec e_\mathrm{n}>0$) as a function depending on the wall temperature $T_\mathrm{wall}$, accommodation coefficients $\alpha_\mathrm{n}, \alpha_\tau$ and possibly also on the incoming velocity $\vec c_\mathrm{i}$. So, we have $R=R(\vec c_\mathrm{r}, \vec c_\mathrm{i})$ which contains $T_\mathrm{wall}, \alpha_\mathrm{n}, \alpha_\tau$ as parameters. $R$ satisfies the normalization condition $\int_{\vec c_\mathrm{r}\cdot\vec e_\mathrm{n}>0}R(\vec c_\mathrm{r}, \vec c_\mathrm{i})\mathrm{d}\vec c_\mathrm{r}=1$ where $R(\vec c_\mathrm{r}, \vec c_\mathrm{i})\mathrm{d}\vec c_\mathrm{r}$ is the probability for the molecule coming at $\vec c_\mathrm{i}$ to reflect into the velocity space element $\mathrm{d}\vec c_\mathrm{r}$ around $\vec c_\mathrm{r}$. The transformation between $f$ at the boundary and $R$ can be completed using the incoming mass flux.

We introduce $f_\mathrm{B}(\vec c)$ as the equivalent distribution function of $f$ observed in $S_\mathrm{local}$ at the reflection point $\vec x_l$ and at the current moment $t$, which means $f_\mathrm{B}(\vec c)=f(t, \vec x_l, \vec c+\vec u_\mathrm{wall})$. After getting the formula of $f_\mathrm{B}(\vec c)$, $F_{l\mathrm{,new}}=f_\mathrm{B}(\vec c_\mathrm{r})$. The distribution $f_\mathrm{B}(\vec c_\mathrm{i})|_{\vec c_\mathrm{i}\cdot\vec e_\mathrm{n}<0}$ of the incoming molecules is known from the molecular information in the adjacent cell. $f_\mathrm{B}(\vec c_\mathrm{r})|_{\vec c_\mathrm{r}\cdot\vec e_\mathrm{n}>0}$ is the distribution of reflecting molecules and related to $R$ as introduced in \cite{Ching2005}:
\begin{equation}\label{eq:kernel}
    f_\mathrm{B}(\vec c_\mathrm{r})(\vec c_\mathrm{r}\cdot\vec e_\mathrm{n})\mathrm{d}\vec c_\mathrm{r}=-\int_{\vec c_\mathrm{i}\cdot\vec e_\mathrm{n}<0}R(\vec c_\mathrm{r}, \vec c_\mathrm{i})f_\mathrm{B}(\vec c_\mathrm{i})(\vec c_\mathrm{i}\cdot\vec e_\mathrm{n})\mathrm{d}\vec c_\mathrm{i}\mathrm{d}\vec c_\mathrm{r}
\end{equation}
Taking integration of Eq. \eqref{eq:kernel} with respect to $\vec c_\mathrm{r}$ over its half velocity space and using the normalization condition of $R(\vec c_\mathrm{r}, \vec c_\mathrm{i})$, we get:
\begin{equation}\label{eq:massbalance}
\begin{aligned}
    &\int_{\vec c_\mathrm{r}\cdot\vec e_\mathrm{n}>0}f_\mathrm{B}(\vec c_\mathrm{r})(\vec c_\mathrm{r}\cdot\vec e_\mathrm{n})\mathrm{d}\vec c_\mathrm{r} \\
    &=-\int_{\vec c_\mathrm{r}\cdot\vec e_\mathrm{n}>0}\int_{\vec c_\mathrm{i}\cdot\vec e_\mathrm{n}<0}R(\vec c_\mathrm{r}, \vec c_\mathrm{i})f_\mathrm{B}(\vec c_\mathrm{i})(\vec c_\mathrm{i}\cdot\vec e_\mathrm{n})\mathrm{d}\vec c_\mathrm{i}\mathrm{d}\vec c_\mathrm{r} \\
    &=-\int_{\vec c_\mathrm{i}\cdot\vec e_\mathrm{n}<0}f_\mathrm{B}(\vec c_\mathrm{i})(\vec c_\mathrm{i}\cdot\vec e_\mathrm{n})\mathrm{d}\vec c_\mathrm{i}
\end{aligned}
\end{equation}
which represents the mass conservation of molecular reflection process.

In the Maxwell diffuse reflection model, $f_\mathrm{B,diffuse}(\vec c_\mathrm{r})=n_\mathrm{eff}(\dfrac{m}{2\pi k_\mathrm{B}T_\mathrm{wall}})^{3/2}$ $\exp(\dfrac{-m\vec c_\mathrm{r}^2}{2k_\mathrm{B}T_\mathrm{wall}})$ where the effective $n_\mathrm{eff}$ will be determined by $f_\mathrm{B}(\vec c_\mathrm{i})$. We assume that $f_\mathrm{B}(\vec c_\mathrm{i})=n_{\mathrm{tr,}k}(\dfrac{m}{2\pi k_\mathrm{B}T_{\mathrm{tr,}k}})^{3/2}\exp[\dfrac{-m(\vec c_\mathrm{i}-(\vec u_{\mathrm{tr,}k}-\vec u_\mathrm{wall}))^2}{2k_\mathrm{B}T_{\mathrm{tr,}k}}]$ where $n_{\mathrm{tr,}k}, \vec u_{\mathrm{tr,}k}, T_{\mathrm{tr,}k}$ are the quantities of cell $k$ close to the reflection point $\vec x_l$. Then, the number $N_\mathrm{in}$ of incoming real molecules on per unit wall surface during per unit time is:
\begin{equation}\label{eq:Num-in}
\begin{aligned}
    N_\mathrm{in}&=-\int_{\vec c_\mathrm{i}\cdot\vec e_\mathrm{n}<0}f_\mathrm{B}(\vec c_\mathrm{i})(\vec c_\mathrm{i}\cdot\vec e_\mathrm{n})\mathrm{d}\vec c_\mathrm{i} \\
    &=n_{\mathrm{tr,}k}\sqrt{\dfrac{k_\mathrm{B}T_{\mathrm{tr,}k}}{2\pi m}}[\exp(-{u'}_\mathrm{in}^2)+\sqrt{\pi}{u'}_\mathrm{in}(1+\mathrm{erf}({u'}_\mathrm{in}))]
\end{aligned}
\end{equation}
where ${u'}_\mathrm{in}=\dfrac{-(\vec u_{\mathrm{tr,}k}-\vec u_\mathrm{wall})\cdot\vec e_\mathrm{n}}{\sqrt{2k_\mathrm{B}T_{\mathrm{tr,}k}/m}}$. Similarly, the number $N_\mathrm{out}$ of reflecting real molecules is:
\begin{equation}\label{eq:Num-out-Maxwell}
\begin{aligned}
    N_\mathrm{out}=\int_{\vec c_\mathrm{r}\cdot\vec e_\mathrm{n}>0}f_\mathrm{B,diffuse}(\vec c_\mathrm{r})(\vec c_\mathrm{r}\cdot\vec e_\mathrm{n})\mathrm{d}\vec c_\mathrm{r}
    =n_\mathrm{eff}\sqrt{\dfrac{k_\mathrm{B}T_\mathrm{wall}}{2\pi m}}
\end{aligned}
\end{equation}
Let $N_\mathrm{out}=N_\mathrm{in}$ as required in Eq. \eqref{eq:massbalance}, we get:
\begin{equation}\label{eq:n-eff-Maxwell}
\begin{aligned}
    n_\mathrm{eff}=n_{\mathrm{tr,}k}\sqrt{\dfrac{T_{\mathrm{tr,}k}}{T_\mathrm{wall}}}[\exp(-{u'}_\mathrm{in}^2)+\sqrt{\pi}{u'}_\mathrm{in}
    (1+\mathrm{erf}({u'}_\mathrm{in}))]
\end{aligned}
\end{equation}
Now, we can compute $F_{l\mathrm{,new}}=f_\mathrm{B,diffuse}(\vec c_\mathrm{r})$ after getting $n_\mathrm{eff}$. We store $n_\mathrm{eff}$ and use it repeatedly for different simulated molecules reflecting on the same subarea close to cell $k$ during the same $\Delta t$ and update $n_\mathrm{eff}$ at the end of each $\Delta t$. Additionally, the scatter kernel $R_\mathrm{diffuse}$ of the Maxwell diffuse reflection model can be determined from $f_\mathrm{B,diffuse}(\vec c_\mathrm{r})$ as we assume that it is independent of the incoming velocity $\vec c_\mathrm{i}$, namely $R_\mathrm{diffuse}=R_\mathrm{diffuse}(\vec c_\mathrm{r})$. Using the formula of $f_\mathrm{B,diffuse}(\vec c_\mathrm{r})$ and Eqs. \eqref{eq:kernel}, \eqref{eq:massbalance}, \eqref{eq:Num-out-Maxwell} we have:
\begin{equation}\label{eq:kernelMaxwell}
\begin{aligned}
    R_\mathrm{diffuse}(\vec c_\mathrm{r})&=\dfrac{f_\mathrm{B,diffuse}(\vec c_\mathrm{r})(\vec c_\mathrm{r}\cdot\vec e_\mathrm{n})}{-\int_{\vec c_\mathrm{i}\cdot\vec e_\mathrm{n}<0}f_\mathrm{B}(\vec c_\mathrm{i})(\vec c_\mathrm{i}\cdot\vec e_\mathrm{n})\mathrm{d}\vec c_\mathrm{i}} \\
    &=\dfrac{f_\mathrm{B,diffuse}(\vec c_\mathrm{r})(\vec c_\mathrm{r}\cdot\vec e_\mathrm{n})}{\int_{\vec c_\mathrm{r}\cdot\vec e_\mathrm{n}>0}f_\mathrm{B,diffuse}(\vec c_\mathrm{r})(\vec c_\mathrm{r}\cdot\vec e_\mathrm{n})\mathrm{d}\vec c_\mathrm{r}} \\
    &=\dfrac{\vec c_\mathrm{r}\cdot\vec e_\mathrm{n}}{2\pi}(\dfrac{m}{k_\mathrm{B}T_\mathrm{wall}})^{2}\exp(\dfrac{-m\vec c_\mathrm{r}^2}{2k_\mathrm{B}T_\mathrm{wall}})
\end{aligned}
\end{equation}
which implies the selecting algorithm of $\vec c_\mathrm{r}$ for the Maxwell diffuse reflection model described in section \ref{sss:updating c_l}.

In the CL reflection model \cite{Carlo1971CLL}, the scatter kernel is the product of three independent parts related respectively to the three components:
\begin{equation}\label{eq:kernelCLL}
\begin{aligned}
    R_\mathrm{CL}(\vec c_\mathrm{r}, \vec c_\mathrm{i})=
    &\dfrac{1}{\sqrt{\pi\alpha_{\tau}}}\exp[\dfrac{-(\tilde{c}_{\mathrm{r,}2}-\sqrt{1-\alpha_\tau}\tilde{c}_{\mathrm{i,}2})^2}{\alpha_\tau}]\times
    \\ &\dfrac{1}{\sqrt{\pi\alpha_{\tau}}}\exp[\dfrac{-(\tilde{c}_{\mathrm{r,}3}-\sqrt{1-\alpha_\tau}\tilde{c}_{\mathrm{i,}3})^2}{\alpha_\tau}]\times
    \\ &\dfrac{\tilde{c}_{\mathrm{r,}1}}{\pi\alpha_\mathrm{n}}\exp[\dfrac{-(\tilde{c}_{\mathrm{r,}1}^2+(1-\alpha_\mathrm{n})
    \tilde{c}_{\mathrm{i,}1}^2)}{\alpha_\mathrm{n}}]\times
    \\ &\int_{0}^{2\pi}\exp[\dfrac{2\sqrt{1-\alpha_\mathrm{n}}\tilde{c}_{\mathrm{r,}1}|\tilde{c}_{\mathrm{i,}1}|}{\alpha_\mathrm{n}}
    \cos\theta]\mathrm{d}\theta
\end{aligned}
\end{equation}
where $|\tilde{c}_{\mathrm{i,}1}|$ is the absolute value of the normalized incoming component $\dfrac{c_{\mathrm{i,}1}}{\sqrt{2k_\mathrm{B}T_\mathrm{wall}/m}}$ where $c_{\mathrm{i,}1}<0$. The selecting algorithm of $\vec c_\mathrm{r}$ was proposed in \cite{Lord1991CLL} based on Eq. \eqref{eq:kernelCLL} and is referred to as CLL reflection model. Again, we assume that $f_\mathrm{B}(\vec c_\mathrm{i})$ is a Maxwell distribution, which is a rough assumption here although it is reasonable when calculating $N_\mathrm{in}$ by Eq. \eqref{eq:Num-in}. Then, $f_\mathrm{B,CL}(\vec c_\mathrm{r})$ is determined from Eqs. \eqref{eq:kernel} and \eqref{eq:kernelCLL}. Unfortunately, it is complicated to calculate $f_\mathrm{B,CL}(\vec c_\mathrm{r})$ by solving the integral of Eq. \eqref{eq:kernel}. A tentative scheme was proposed in \cite{Jun2011RGD} to simplify the calculation. Note that the major differences between $f_\mathrm{B,diffuse}(\vec c_\mathrm{r})$ and $R_\mathrm{diffuse}$ are that the former contains a parameter $n_\mathrm{eff}$ but the later contains $\vec c_\mathrm{r}\cdot\vec e_\mathrm{n}$. So, $\tilde{c}_{\mathrm{r,}1}$ is removed from $R_\mathrm{CL}$ and a new parameter $a$ is added to describe $f_\mathrm{B,CL}(\vec c_\mathrm{r})$.  The mass conservation principle of Eq. \eqref{eq:massbalance} is used to determine the parameter $a$. Consequently, the tentative formula of $f_\mathrm{B,CL}$ depends not only on $\vec c_\mathrm{r}$ but also on $\vec c_\mathrm{i}$, which is inconsistent with the definition of Eq. \eqref{eq:kernel} where $f_\mathrm{B,CL}$ is independent of $\vec c_\mathrm{i}$. Some simulation results show that the tentative formula of $f_\mathrm{B,CL}$ is useful when $\alpha_\tau, \alpha_\mathrm{n}$ are very close to $1$ ($\alpha_\tau=\alpha_\mathrm{n}=0.98$) \cite{Jun2011ICNMM}. This is because the tentative formula of $f_\mathrm{B,CL}$ can degenerate to the correct $f_\mathrm{B,diffuse}$ when $\alpha_\tau=\alpha_\mathrm{n}=1$ and so its error is negligible when $\alpha_\tau, \alpha_\mathrm{n}\to1$.

In the specular reflection model, $\vec c_\mathrm{r}\equiv\vec c_\mathrm{i}-2\vec e_\mathrm{n}(\vec c_\mathrm{i}\cdot\vec e_\mathrm{n})$ and so $R_\mathrm{specular}=\delta(\vec c_\mathrm{r}-(\vec c_\mathrm{i}-2\vec e_\mathrm{n}(\vec c_\mathrm{i}\cdot\vec e_\mathrm{n})))$. Submitting $R_\mathrm{specular}$ into Eq. \eqref{eq:kernel}, $f_\mathrm{B,specular}(\vec c_\mathrm{r})=f_\mathrm{B}(\vec c_\mathrm{i})$, which implies $F_{l\mathrm{,new}}=F_l$ as $F_{l\mathrm{,new}}=f_\mathrm{B,specular}(\vec c_\mathrm{r})$ and $f_\mathrm{B}(\vec c_\mathrm{i})=f(t, \vec x_l, \vec c_l)$ and $F_l$ is equal to $f(t, \vec x_l, \vec c_l)$ before reflecting.
\subsection{Calculation of flux on boundary}\label{ss:algorithm summary}
As in the DSMC method, it is convenient for the DSBGK method to calculate the flux $\Gamma(Q)$ of any molecular quantity $Q(\vec c)$ in unit time and across unit area of the boundary surface:
\begin{equation}\label{eq:flux}
    \Gamma(Q)=\dfrac{1}{\Delta t\Delta S}\sum_lN_l[Q(\vec c_\mathrm{i})-Q(\vec c_\mathrm{r})]_l
\end{equation}
where the summation is over all those simulated molecules reflecting on the subarea $\Delta S$ during the time step $\Delta t$, $Q(\vec c_\mathrm{i})$ and $Q(\vec c_\mathrm{r})$ are the incoming and reflecting quantities, respectively. Let $Q=m\vec c$ and $m\vec c^2/2$ and then $\Gamma(Q)$ represents the stress and heat flux, respectively.
\subsection{Summary of the DSBGK algorithm}\label{ss:algorithm summary}
1. Initialization. Generate many cells and simulated molecules and assign them with initial values for $n_{\mathrm{tr,}k}, \vec u_{\mathrm{tr,}k}, T_{\mathrm{tr,}k}$ and $\vec x_l, \vec c_l, F_l, N_l$, respectively.

2. Each simulated molecule moves uniformly and in a straight line before encountering boundary. During each $\Delta t$, the trajectory of any particular molecule $l$ may be divided into several segments (see Fig. \ref{fig:schematic model}). Then, $\vec x_l, F_l, N_l$ are updated \textit{deterministically} along each segment in sequence. When encountering the wall boundary, $\vec c_l$ is updated randomly according to the reflection model and then $F_l$ is updated correspondingly. In open problems, simulated molecules are removed from the flow domain when moving across the open boundary during each $\Delta t$ and new simulated molecules are generated at the open boundary at the end of each $\Delta t$. The variables $n_{\mathrm{tr,}k}, \vec u_{\mathrm{tr,}k}, T_{\mathrm{tr,}k}$ of each cell $k$ is updated at the end of each $\Delta t$.

3. After convergence, $n_{\mathrm{tr,}k}, \vec u_{\mathrm{tr,}k}, T_{\mathrm{tr,}k}$ are used as the discrete solutions of $n, \vec u, T$ at each cell $k$.
\section{Simulation Results}\label{s:simulation}
In DSBGK simulations, the parameter $\upsilon$ is selected to satisfy the coefficient of viscosity $\mu$ or heat conduction $\kappa$ by Eq. \eqref{eq:upsilon}. For problems where the momentum exchange is the dominant effect, we use $\upsilon=\upsilon(\mu)\equiv nk_\mathrm{B}T/\mu$ to satisfy $\mu$. For problems where the heat conduction is the dominant effect, we select $\upsilon=\upsilon(\kappa)\equiv5nk_\mathrm{B}^2T/(2m\kappa)$ to satisfy $\kappa$. Note that $\mu$ is given usually. For monoatomic gas where the Prandtl number is $Pr=2/3$ and the specific heat capacity at constant pressure is $C_p=5k_\mathrm{B}/(2m)$, we have $\upsilon(\kappa)=2nk_\mathrm{B}T/(3\mu)$ as $\kappa=C_p\mu/Pr$.
\subsection{Lid-driven cavity flow}\label{ss:Cavity}
\begin{figure}[H] % 'H' or '!htb'
\centering
\includegraphics[width=0.45\textwidth]{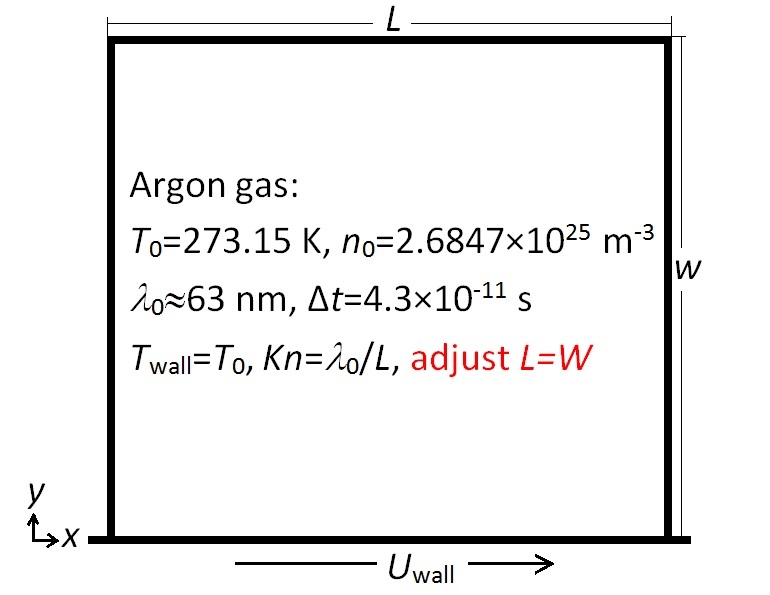}
\caption{Schematic model of lid-driven cavity flow.}
\label{fig:CavityModel}
\end{figure}

 The results were reported first in \cite{Jun2011RGD}. The sizes $L=W$ are regulated to change the $Kn$ number. The Maxwell boundary condition is used and the cell number is $20\times20$ for $Kn=0.063$ and 6.3. We set $\upsilon=\upsilon(\mu)$ in the DSBGK simulations. In order to reduce the influence due to fluctuation in the number density distribution observed in the DSBGK simulations of closed problems (see the following Fig. \ref{fig:CavityflowKn6.3U20}), the number of simulated molecules per cell is about 2000 in the DSBGK simulations at $Kn=0.063$ and 6.3.

\begin{figure}[H]
  \centering
  \includegraphics[width=0.45\textwidth]{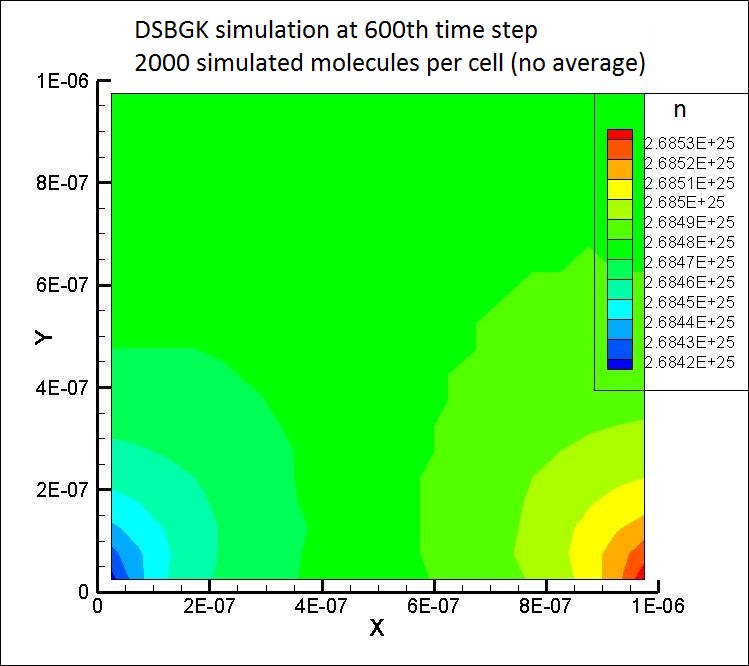}
  \includegraphics[width=0.45\textwidth]{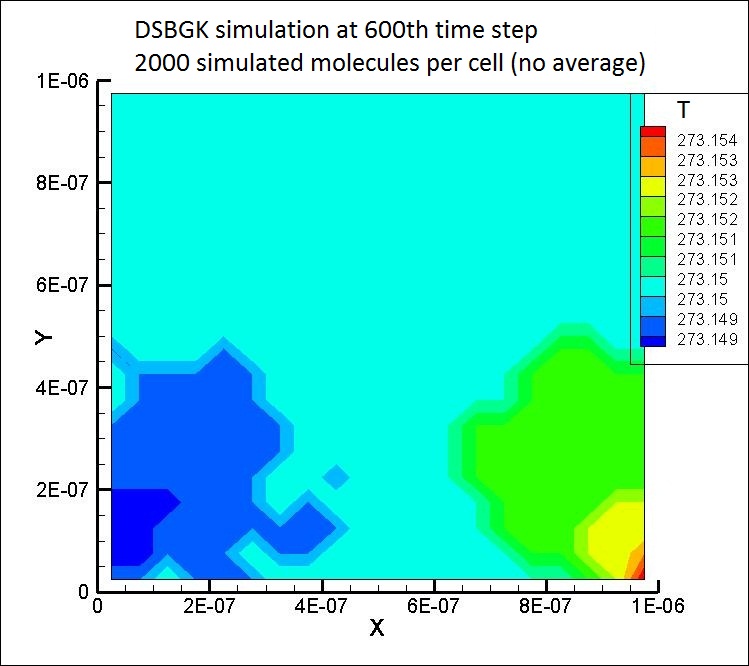} \\
  \includegraphics[width=0.45\textwidth]{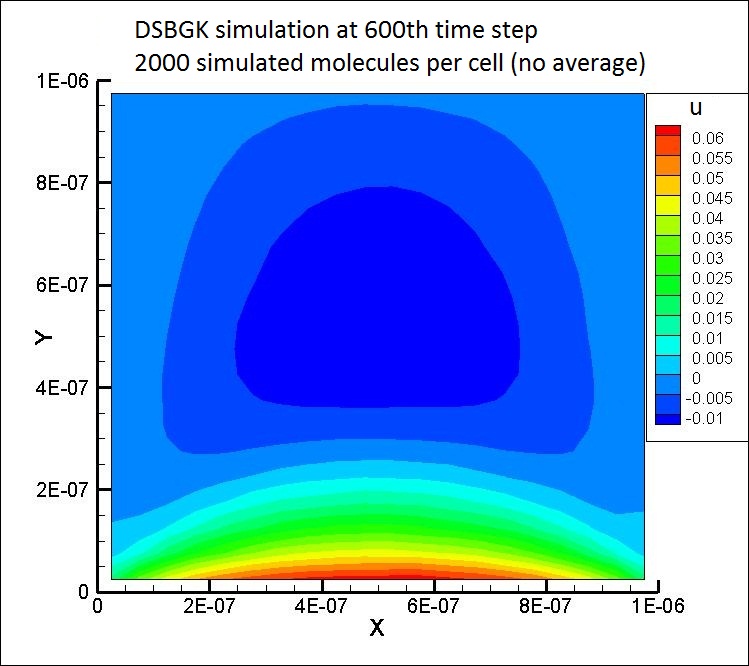}
  \includegraphics[width=0.45\textwidth]{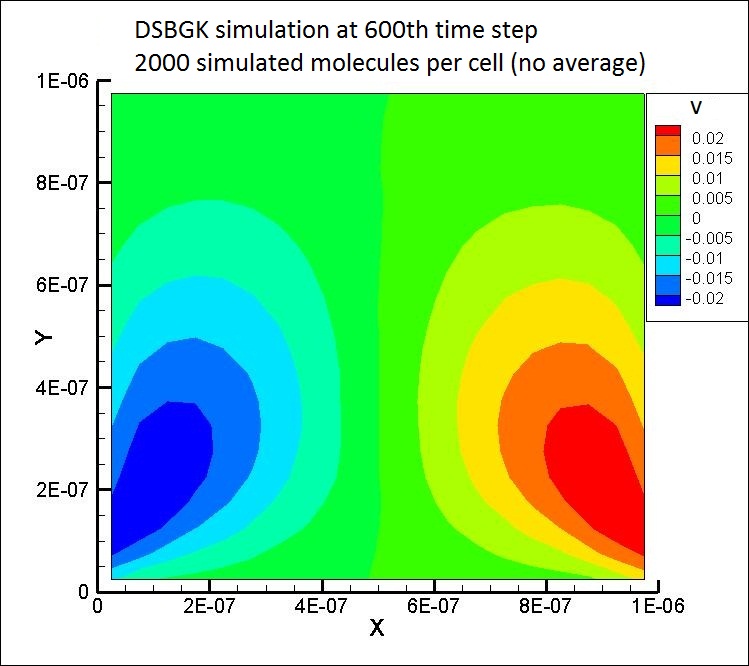}
  \caption{Transient results of DSBGK simulation of lid-driven problem, $Kn=0.063$ and $U_\mathrm{wall}$=0.1 m/s, 7 minutes of CPU time.}
  \label{fig:CavityflowDSBGK}
\end{figure}

To show the high efficiency of DSBGK simulations at low velocity, we choose a very small driven velocity $U_\mathrm{wall}=0.1$ m/s. Fig. \ref{fig:CavityflowDSBGK} shows the transient results (no average) of DSBGK simulation at 600$^{th} \Delta t$ taking about 7 minutes of computational time of one CPU on Lenovo laptop E43A. We can output many transient results at different moments of interest at the additional cost of negligible computational time which is used for writing data to the hard disc. From the efficiency point of view, the DSBGK method is a promising tool for studying transient problems. But, the time coordinate in DSBGK simulations is not synchronous with the real time in physical problems due to the hysteresis effect \cite{Jun2011ICNMM} of DSBGK simulations which are based on the auto-regulation schemes of Eq. \eqref{eq:auto}. New techniques, like time rescaling, are required to reduce the magnitude of hysteresis effect.

\begin{figure}[H]
  \centering
  \includegraphics[width=0.45\textwidth]{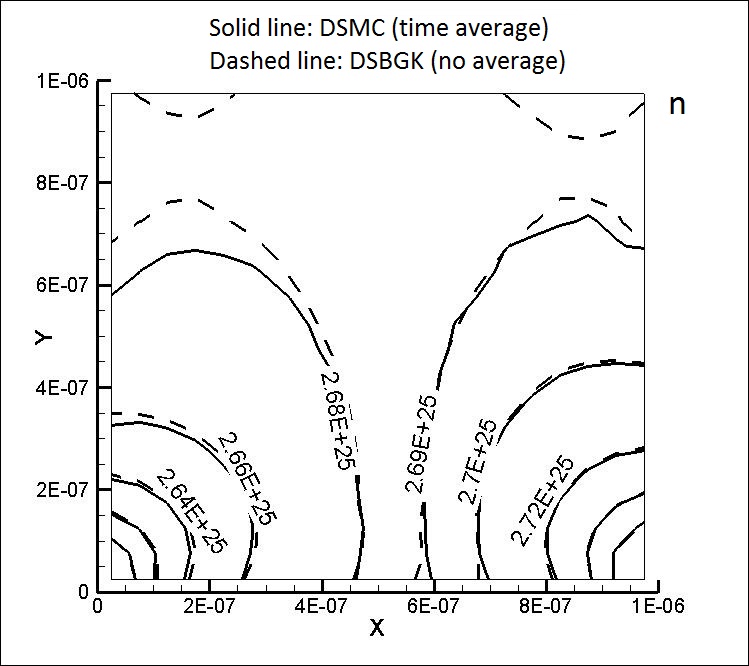}
  \includegraphics[width=0.45\textwidth]{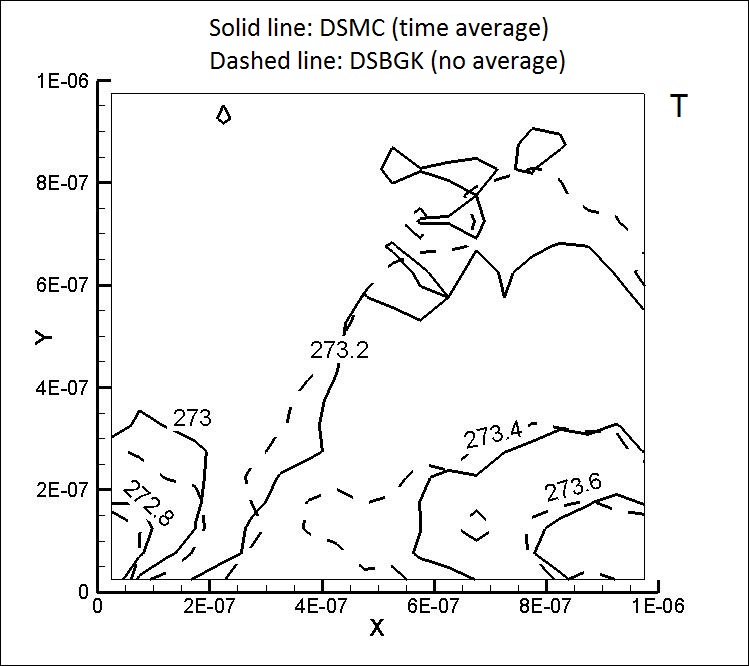} \\
  \includegraphics[width=0.45\textwidth]{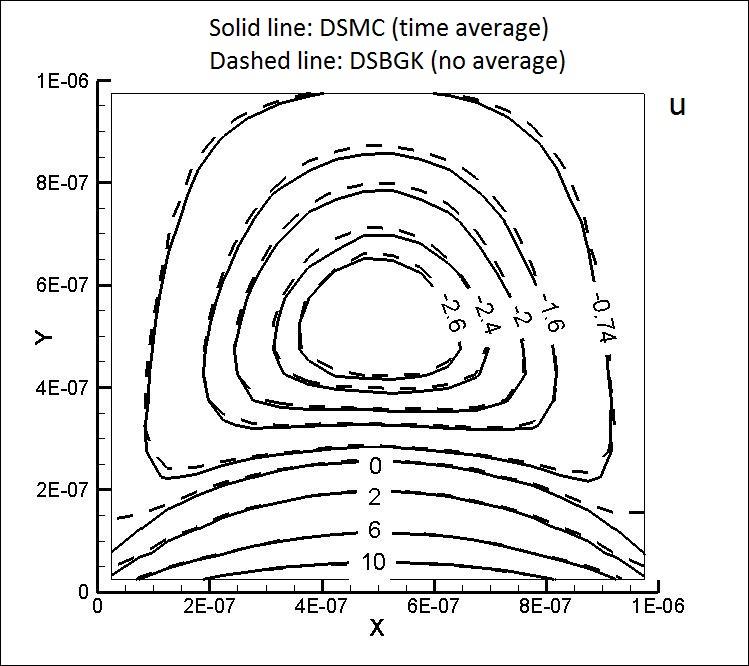}
  \includegraphics[width=0.45\textwidth]{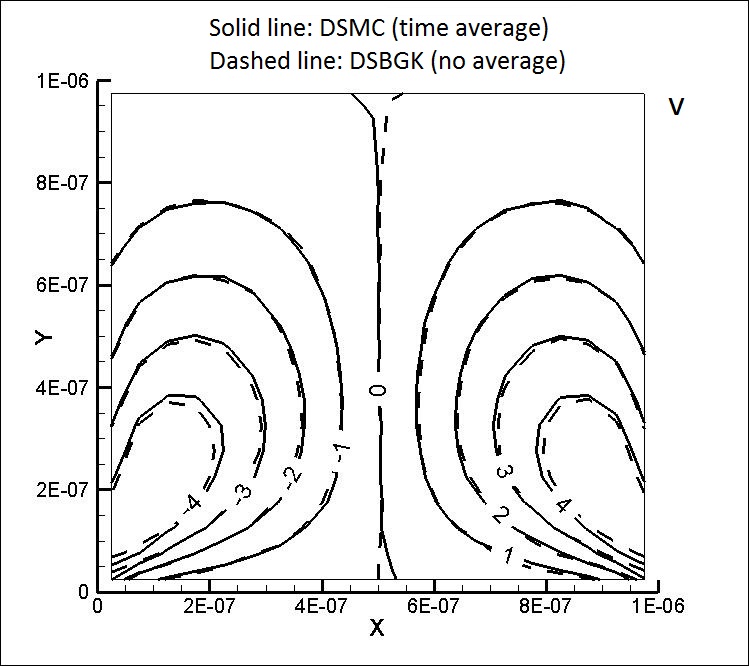}
  \caption{Comparison between DSMC and DSBGK methods in lid-driven problem, $Kn=0.063$ and $U_\mathrm{wall}$=20 m/s.}
  \label{fig:CavityflowKn0.063U20}
\end{figure}

The driven velocity $U_\mathrm{wall}$ increases to 20 m/s and the transient DSBGK results at 600$^{th} \Delta t$ are given in Fig. \ref{fig:CavityflowKn0.063U20} with verification by the DSMC results. The DSBGK simulation uses about 7 minutes again, which implies that the computational time used by the DSBGK simulation is almost independent of the magnitude of deviation from equilibrium state as the average process is avoided here. The DSMC simulation takes about 30 hours using 67 molecules per cell and about $3.4\times10^5$ samples (sampling once every $4 \Delta t$). The computational time required by DSMC simulation for the above case of $U_\mathrm{wall}=0.1$ m/s can be estimated by considering the fact that the computational time of DSMC simulation is almost inversely proportional to the square of Mach number, roughly $30\times200^2$ hours.

\begin{figure}[H]
  \centering
  \includegraphics[width=0.45\textwidth]{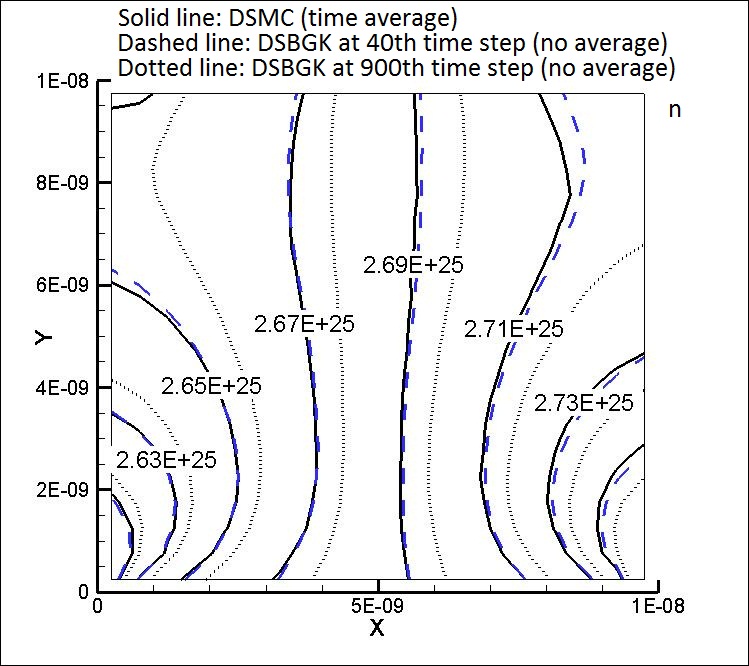}
  \includegraphics[width=0.45\textwidth]{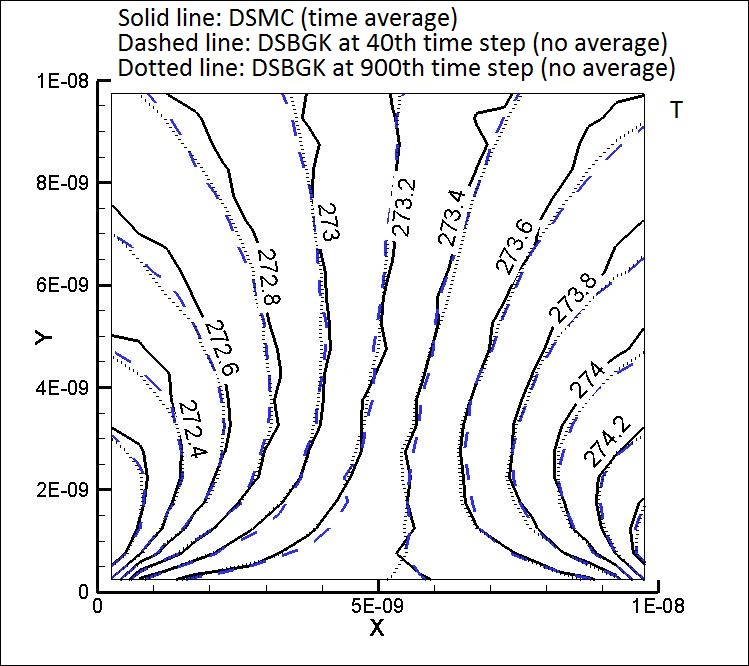} \\
  \includegraphics[width=0.45\textwidth]{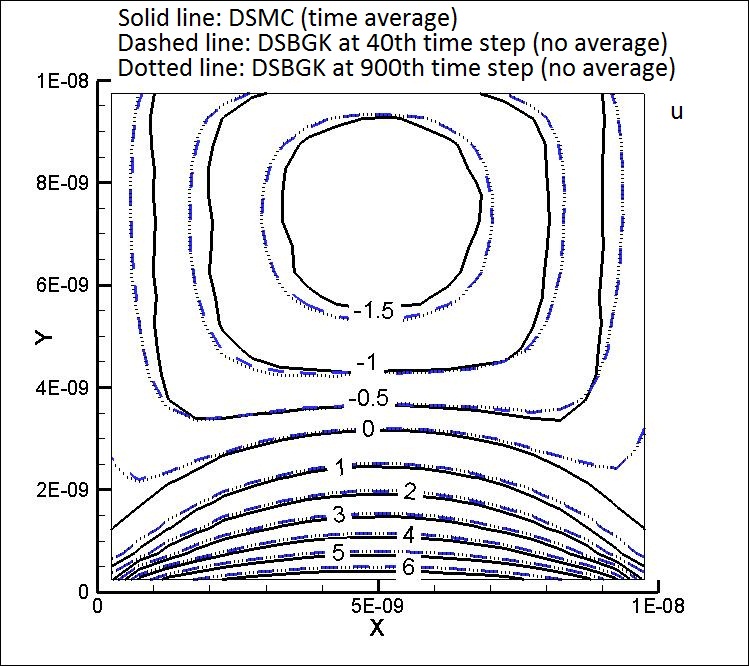}
  \includegraphics[width=0.45\textwidth]{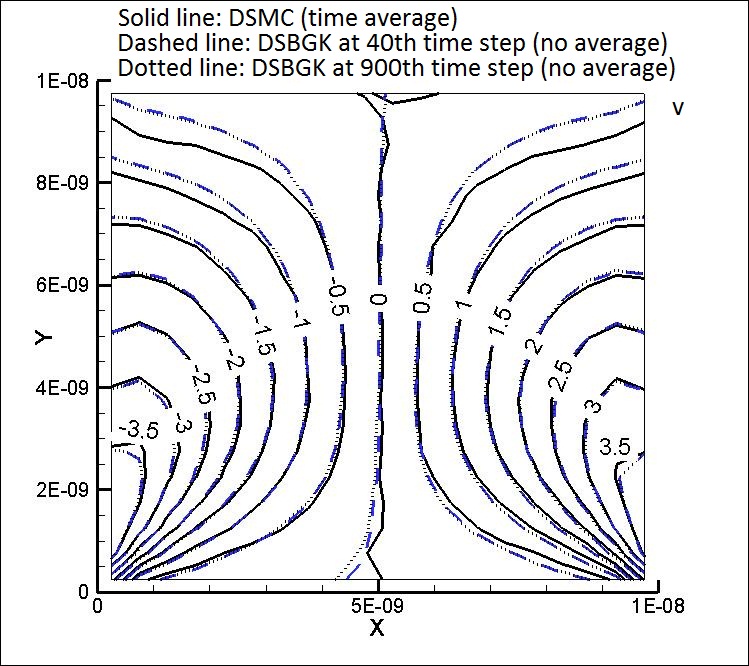}
  \caption{Comparison between DSMC and DSBGK methods in lid-driven problem, $Kn=6.3$ and $U_\mathrm{wall}$=20 m/s.}
  \label{fig:CavityflowKn6.3U20}
\end{figure}

We set $U_\mathrm{wall}=20$ m/s and increase $Kn$ to 6.3. The DSBGK transient results at $40^{th} \Delta t$ agree very well with the DSMC results. The DSBGK distributions of $T, u, v$ remain unchanged after $40 \Delta t$. But, the DSBGK distribution of $n$ can not stay at steady state and its deviation at $900^{th} \Delta t$ from the DSMC result is remarkable. This drawback of the DSBGK method in \textit{closed} problems implies that the ensemble-average process (if necessary) should be used for quantities related to $n$. In \textit{open} problems, the unphysical fluctuation of $n$ is eliminated by the fixed $n$ at open boundaries and so the more-efficient time-average process can be used (see the results of channel flow). The DSBGK simulation within $40 \Delta t$ takes about 11 minutes and the computational time for each $\Delta t$ is increased compared to that of $Kn=0.063$, which is because the molecular reflection on the wall becomes frequent and more computational time is used to generate random fractions.

\subsection{Couette flow}\label{ss:Couette}
The same results were reported in \cite{Jun2011ICNMM}. The distance $L$ between two plates is regulated to change $Kn$. The cell number is 200, 20, 20 for $Kn=0.01$, 0.1, 1, respectively. The Maxwell boundary condition is used. $\upsilon=\upsilon(\mu)$ and each cell contains about 2550 simulated molecules in the DSBGK simulations.

\begin{figure}[H]
  \centering
  \includegraphics[width=0.45\textwidth]{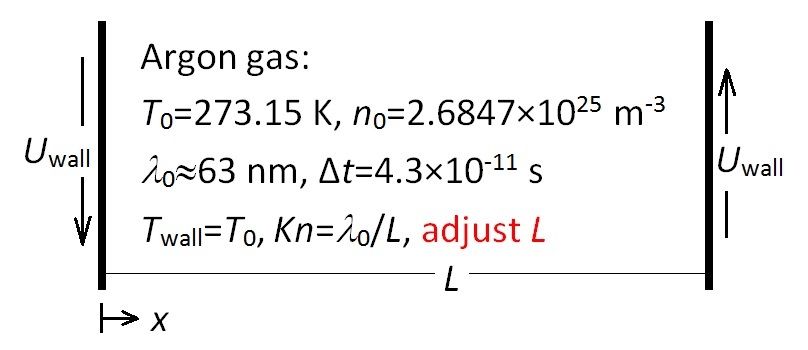}
  \includegraphics[width=0.45\textwidth]{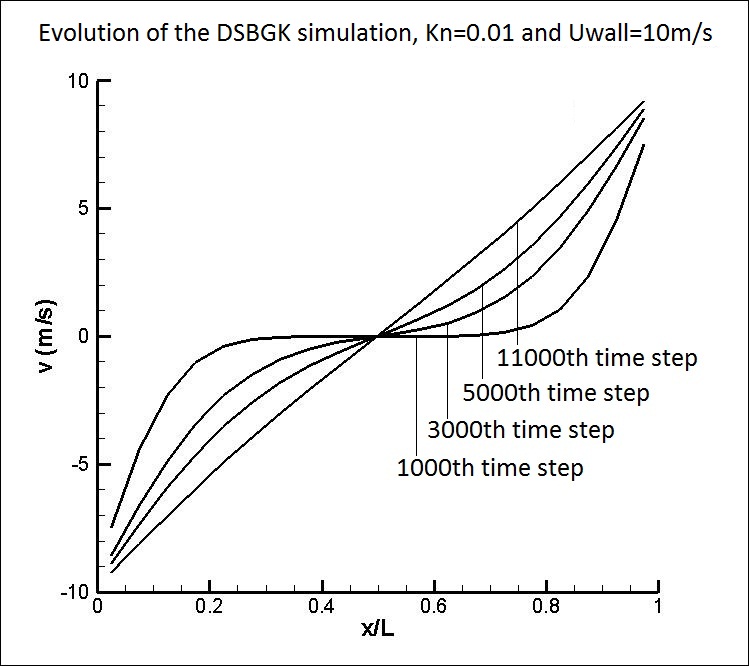} \\
  \includegraphics[width=0.45\textwidth]{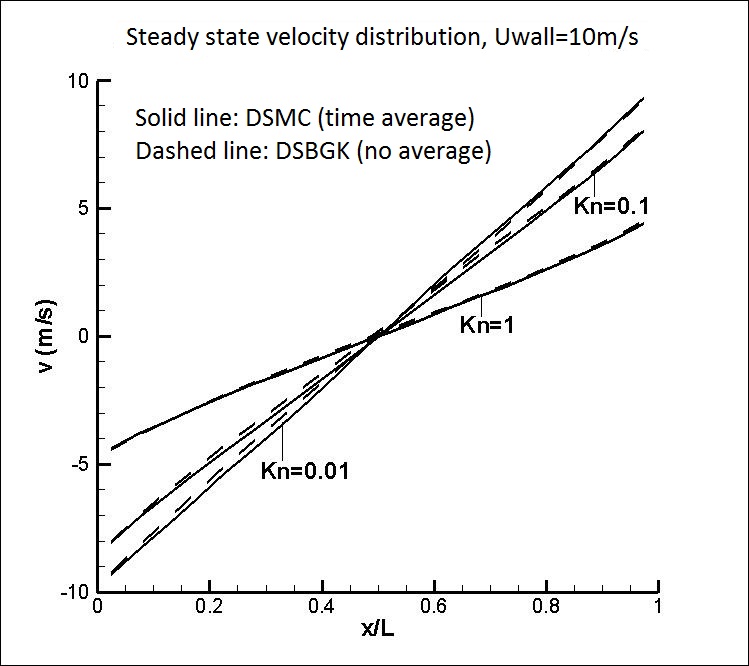}
  \includegraphics[width=0.45\textwidth]{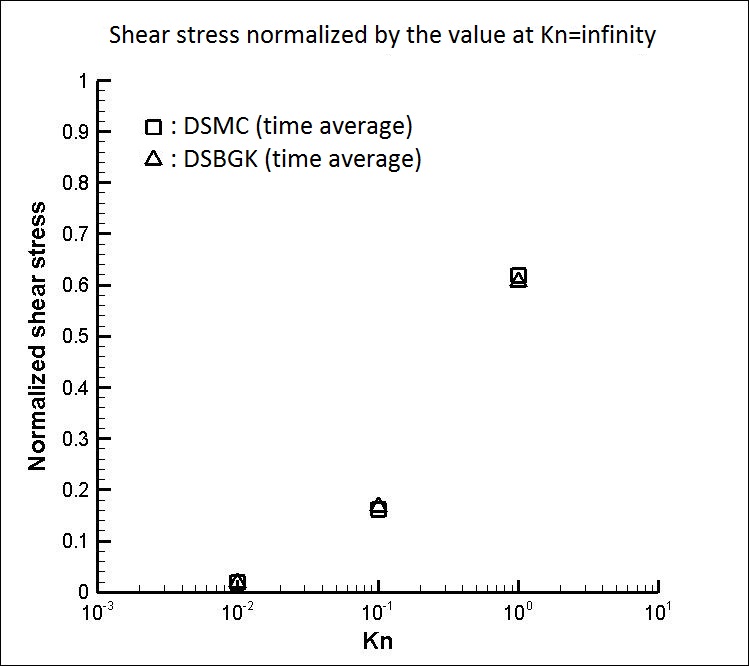}
  \caption{Comparison between DSMC and DSBGK methods in Couette flow problem \cite{Jun2011ICNMM}.}
  \label{fig:Couetteflow}
\end{figure}
\subsection{Thermal transpiration flow}\label{ss:Thermal}
\begin{figure}[H]
\centering
\includegraphics[width=0.45\textwidth]{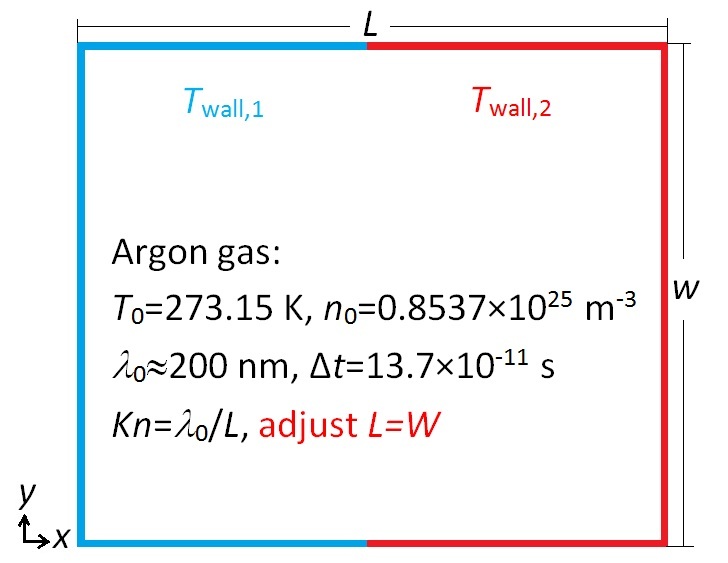}
\caption{Schematic model of thermal transpiration flow.}
\label{fig:ThermalModel}
\end{figure}

This problem was studied first in \cite{Kazuo2001Thermal} where $T_{\mathrm{wall,}2}/T_{\mathrm{wall,}1}$=2. We set $T_{\mathrm{wall,}2}/T_{\mathrm{wall,}1}=1.05T_0/0.95T_0\approx1.105$ to show the high efficiency of DSBGK simulations at low velocity. The sizes $L=W$ are regulated to change $Kn$. The cell number is $40\times40$ for $Kn=0.2$ and the Maxwell boundary condition is used. Each cell contains about 1000 simulated molecules and $\upsilon=\upsilon(\kappa)$ in DSBGK simulation as the heat conduction is the dominant effect.

\begin{figure}[H]
  \centering
  \includegraphics[width=0.45\textwidth]{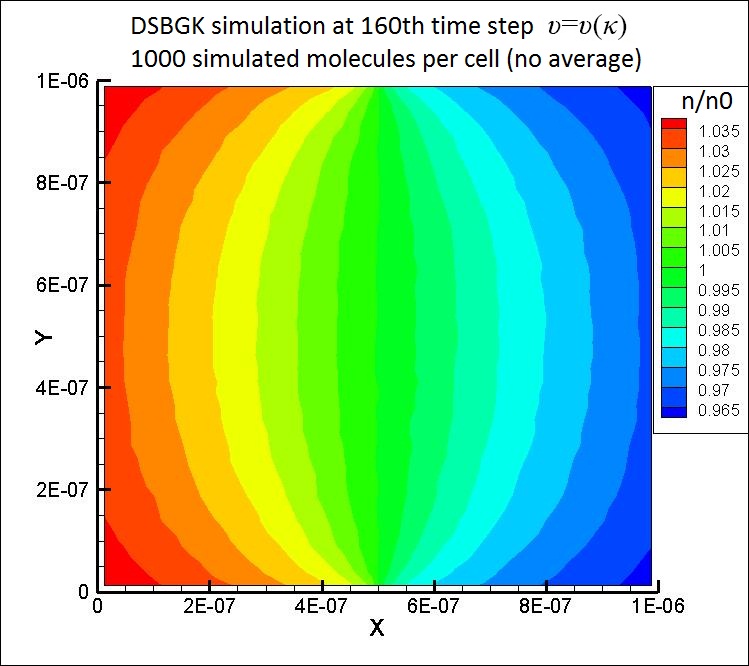}
  \includegraphics[width=0.45\textwidth]{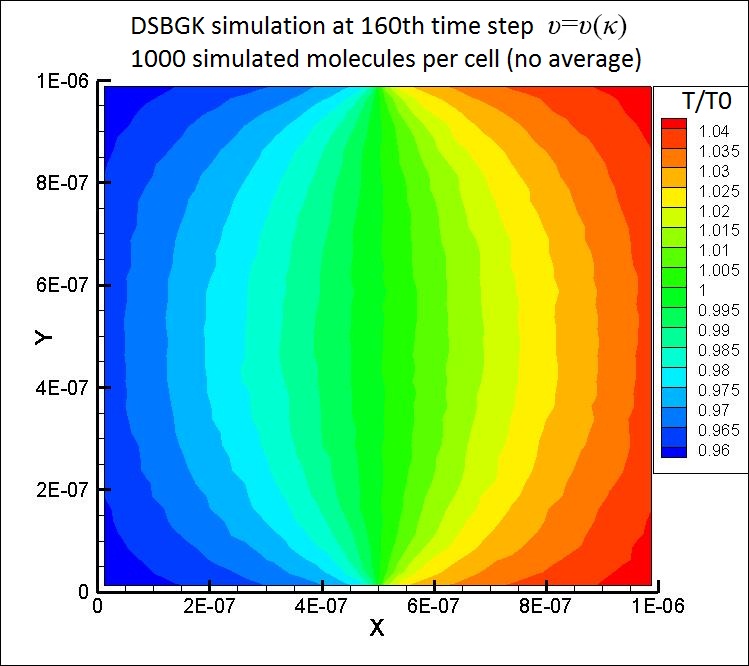} \\
  \includegraphics[width=0.45\textwidth]{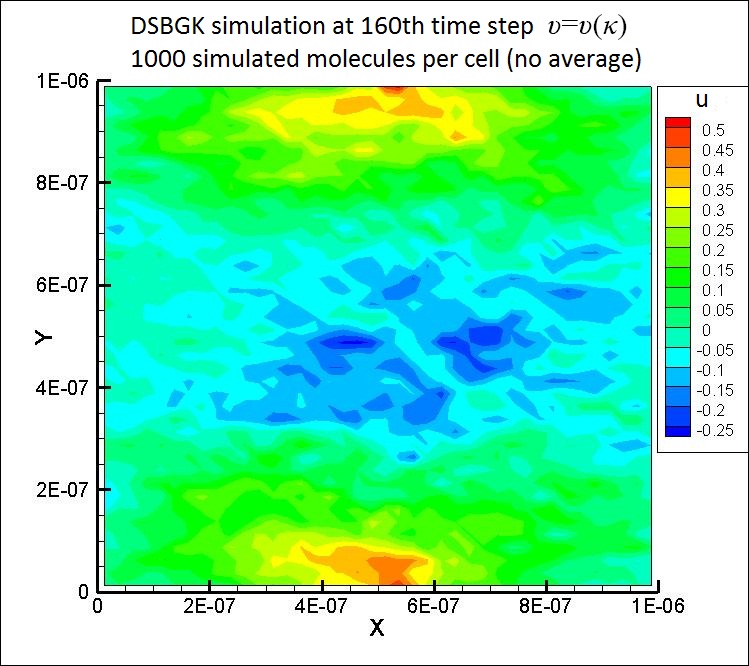}
  \includegraphics[width=0.45\textwidth]{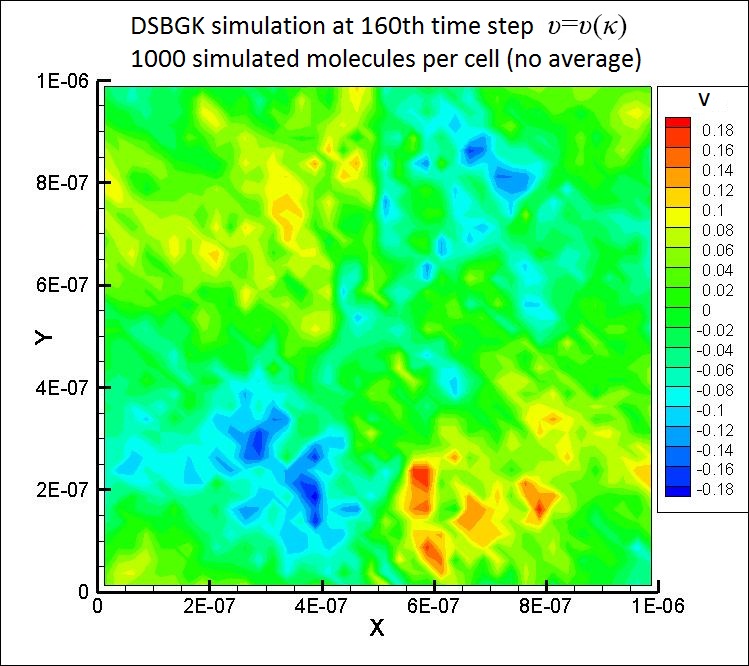} \\
  \caption{Transient results of DSBGK simulation of thermal transpiration problem, $Kn=0.2$, $T_{\mathrm{wall,}1}$=0.95$T_0$, $T_{\mathrm{wall,}2}$=1.05$T_0$, 8 minutes of CPU time.}
  \label{fig:ThermalflowDSBGK}
\end{figure}

\begin{figure}[H]
  \centering
  \includegraphics[width=0.45\textwidth]{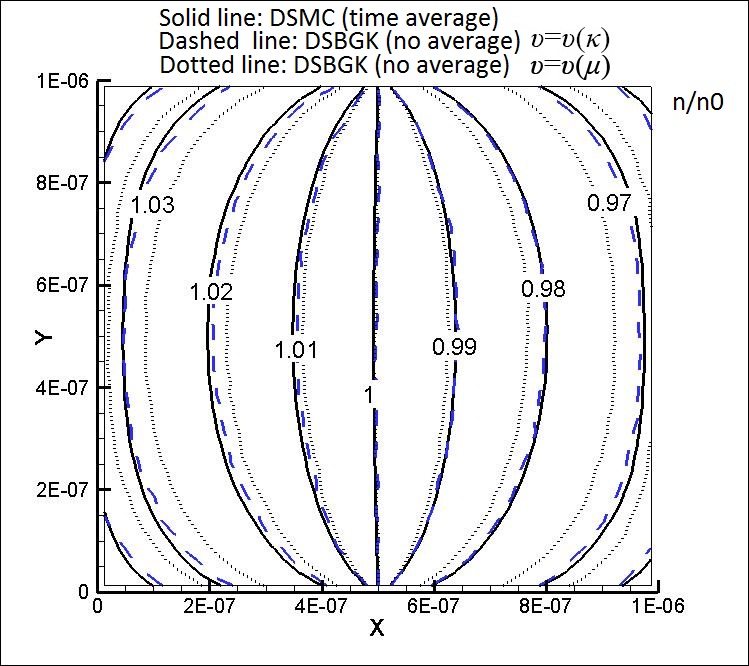}
  \includegraphics[width=0.45\textwidth]{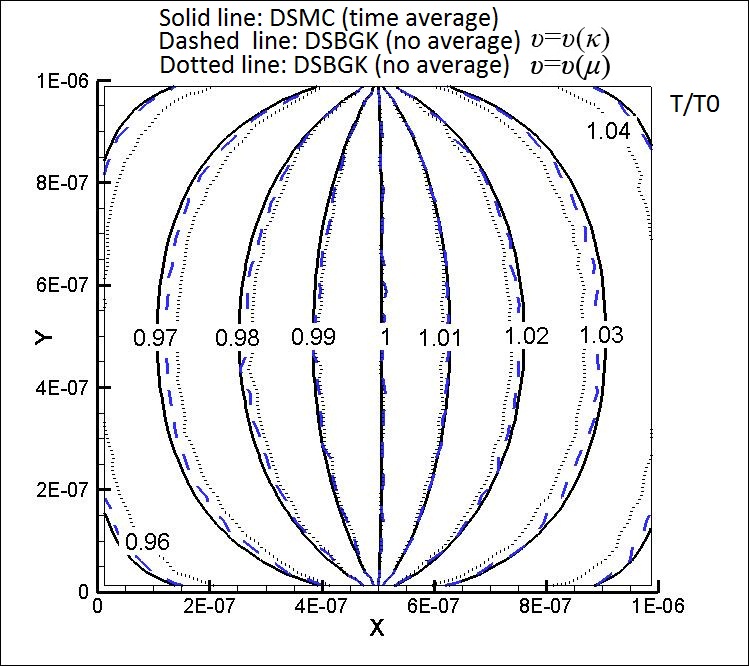} \\
  \includegraphics[width=0.45\textwidth]{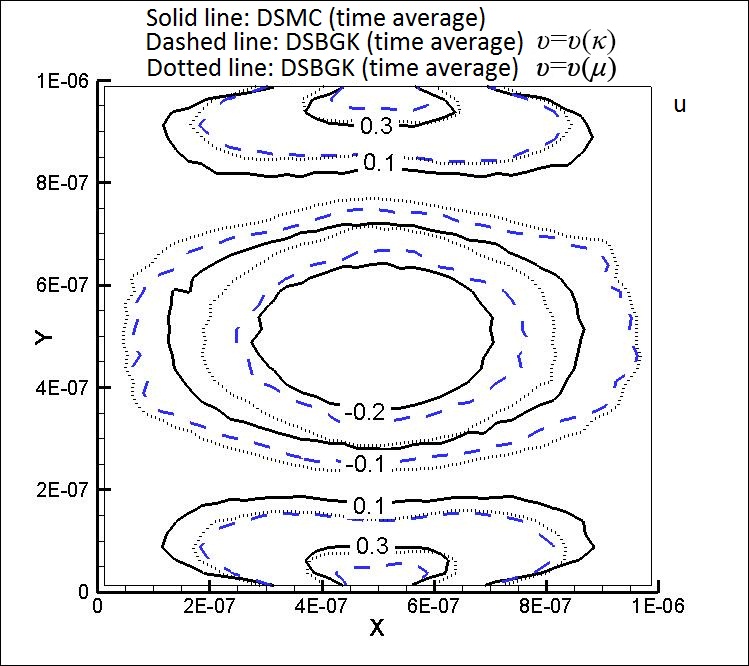}
  \includegraphics[width=0.45\textwidth]{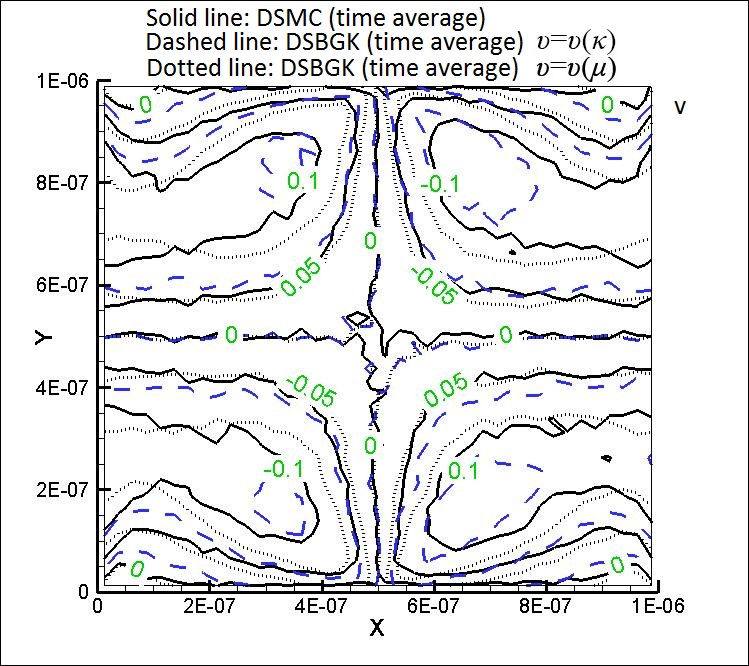} \\
  \includegraphics[width=0.45\textwidth]{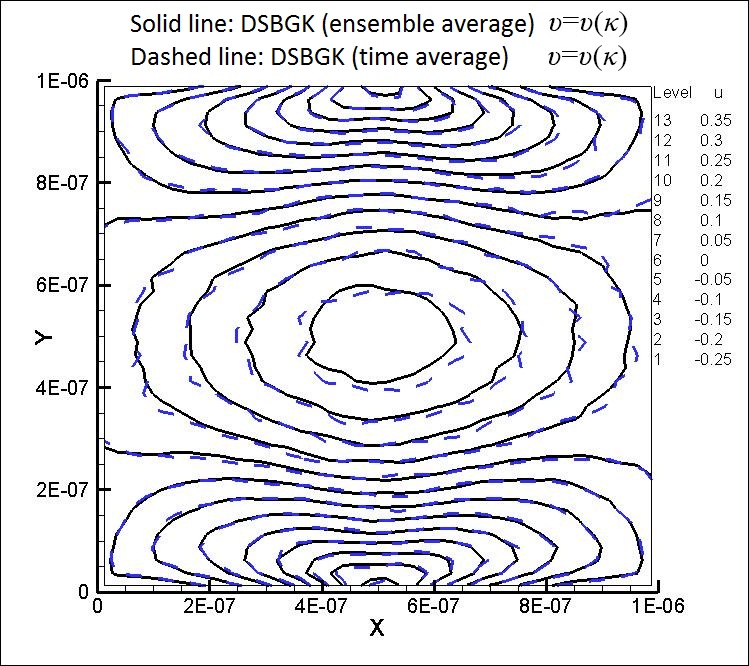}
  \includegraphics[width=0.45\textwidth]{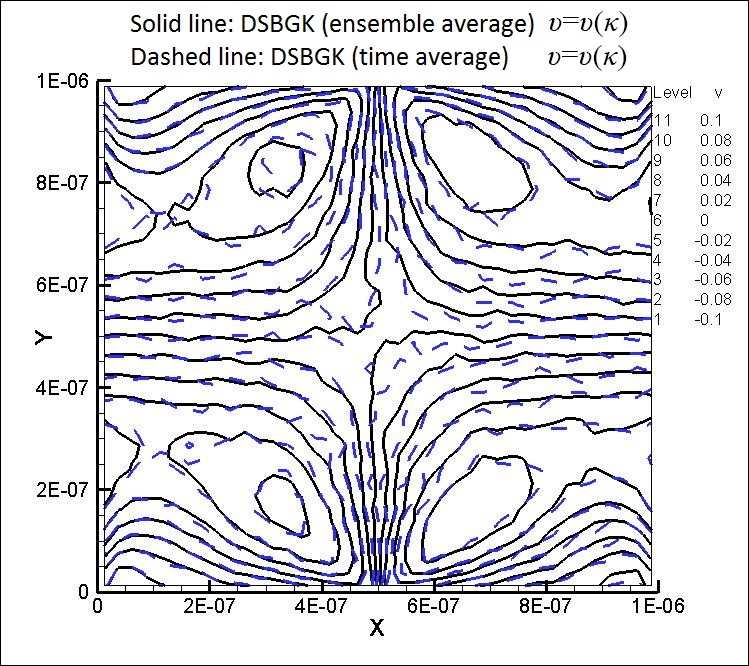} \\
  \caption{Comparison between DSMC and DSBGK methods in thermal transpiration problem, $Kn=0.2$, $T_{\mathrm{wall,}1}$=0.95$T_0$ and $T_{\mathrm{wall,}2}$=1.05$T_0$.}
  \label{fig:ThermalflowKn0.2}
\end{figure}

The DSBGK simulation converges after 160 $\Delta t$ taking about 8 minutes of computational time. The transient DSBGK results are given in Fig. \ref{fig:ThermalflowDSBGK}. The transient $n$ and $T$ are smooth but the transient $u$ and $v$ contain large stochastic noise, which is because that the variation of $T$ is the active factor and has strong correlation with the variation of $n$ through the rough balance of pressure. However, the variations of $u$ and $v$ are passive factors and so sensitive to stochastic noise. In order to present smooth results of $u$ and $v$ for clear verification, we use the time-average process after $160 \Delta t$ to reduce noise and collect 1500 samples (sampling at each $\Delta t$) making the total computational time about 79 minutes. The DSBGK smooth results are given in Fig. \ref{fig:ThermalflowKn0.2} with comparison by the DSMC time-average results. The DSBGK results using $\upsilon=\upsilon(\mu)$ are given together to show the dependence of $\upsilon$ on different problems. The comparison shows that we should select $\upsilon=\upsilon(\kappa)$ in the thermal transpiration problem. In addition, Fig. \ref{fig:ThermalflowKn0.2} shows the agreement between the ensemble-average and time-average for sampling $u$ and $v$ in the DSBGK simulation, which is consistent with the conclusion of Fig. \ref{fig:CavityflowKn6.3U20} that the time-average process is valid for sampling $u$ and $v$.

\begin{figure}[H]
  \centering
  \includegraphics[width=0.45\textwidth]{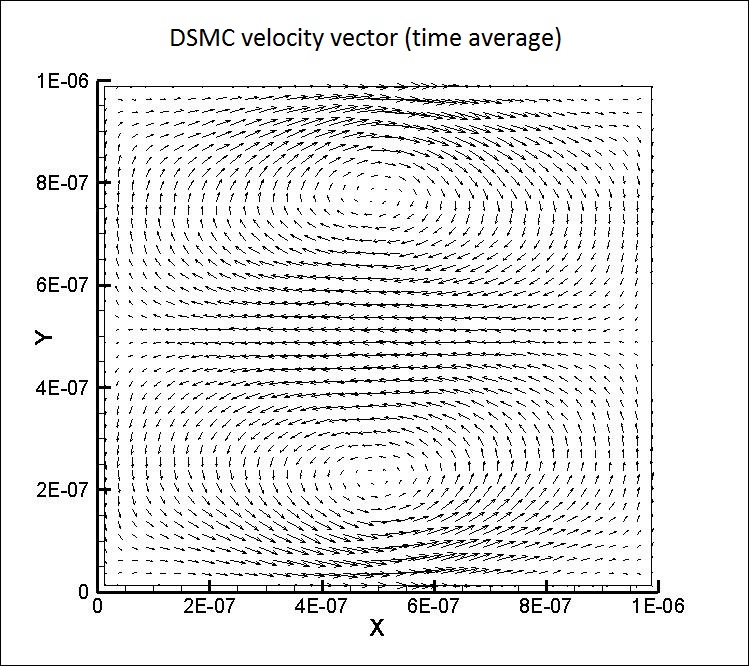}
  \includegraphics[width=0.45\textwidth]{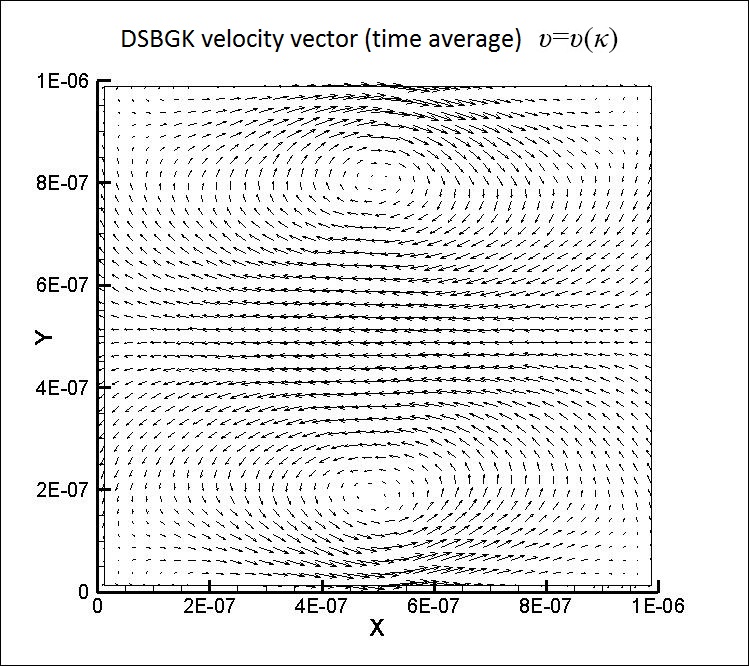} \\
\end{figure}

\subsection{Channel flow}\label{ss:Channel}
\begin{figure}[H]
\centering
\includegraphics[width=0.53\textwidth]{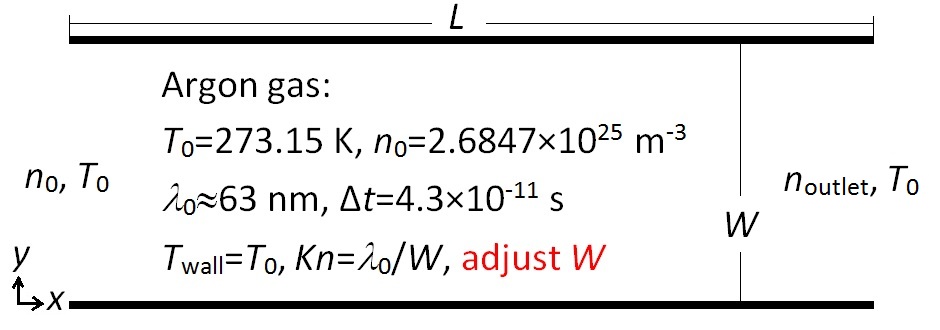}
\caption{Schematic model of channel flow.}
\label{fig:ChannelModel}
\end{figure}

 \begin{figure}[H]
  \centering
  \includegraphics[width=0.45\textwidth]{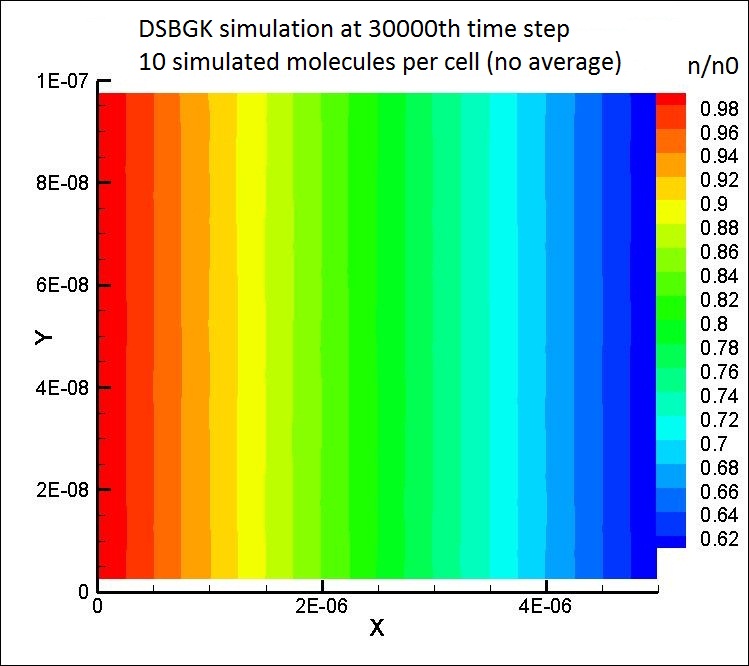}
  \includegraphics[width=0.45\textwidth]{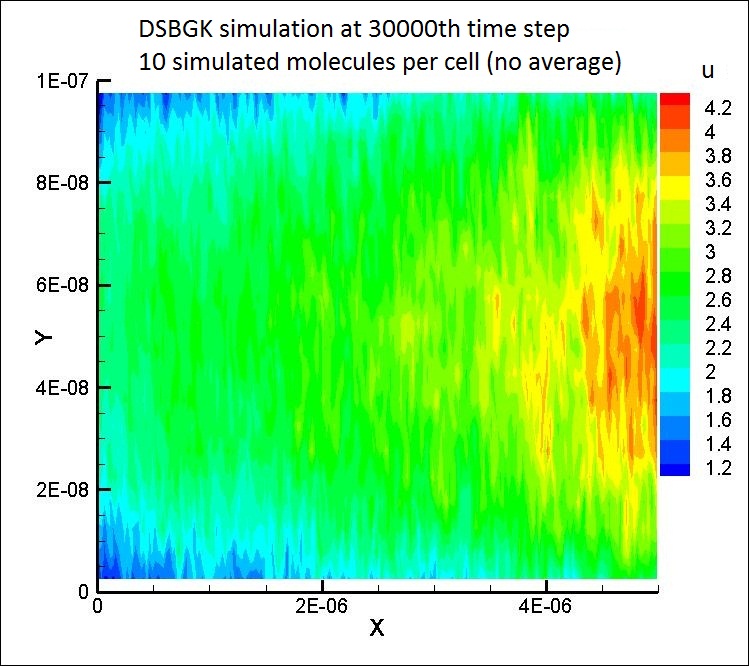} \\
  \includegraphics[width=0.45\textwidth]{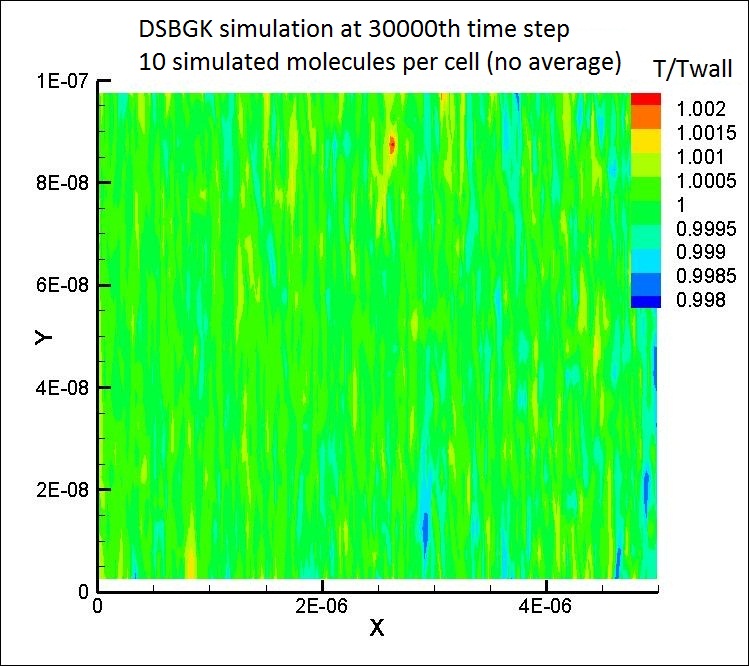}
  \includegraphics[width=0.45\textwidth]{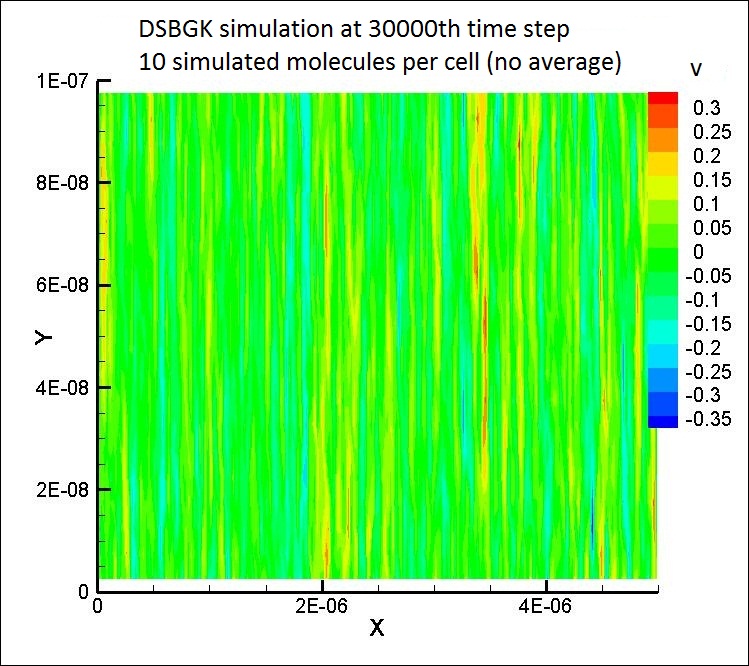} \\
  \caption{Transient results of DSBGK simulation of channel flow problem, $Kn=0.63$ and $n_\mathrm{outlet}$=0.6$n_0$, 36 minutes of CPU time.}
  \label{fig:ChannelflowDSBGK}
\end{figure}

The DSBGK simulations of channel flows driven by pressure difference were reported first in \cite{Jun2011ICNMM}. Here, $L=$ 5 microns and $W$ is regulated to change $Kn$. The cell number is $200\times20$ for $Kn=0.63$ and the Maxwell boundary condition is used. We set $\upsilon=\upsilon(\mu)$ in the DSBGK simulations. To show the stability improvement of DSBGK simulations in open problems, we appropriately choose the initial value $N_{l,t=0}$ of $N_l$ of all simulated molecules such that the number of simulated molecules per cell is about 10 at the initial state. The number density at the outlet is $0.6n_0$ and equal to the initial value $n_0$ at the inlet. In order to make the number of simulated molecules per cell almost uniform during the simulation process, the initial value of $N_l$ for the new simulated molecules at the inlet is larger than that at the outlet and the ratio is $N_{l\mathrm{,init,inlet}}/N_{l\mathrm{,init,outlet}}=n_{\mathrm{inlet}}/n_{\mathrm{outlet}}=1/0.6$. Specifically, we set $N_{l\mathrm{,init,inlet}}=N_{l,t=0}n_{\mathrm{inlet}}/n_0=N_{l,t=0}$ and $N_{l\mathrm{,init,outlet}}=N_{l,t=0}n_{\mathrm{outlet}}/n_0=0.6N_{l,t=0}$ to maintain the number of simulated molecules per cell approximately equal to 10 during the simulation process.

\begin{figure}[H]
  \centering
  \includegraphics[width=0.45\textwidth]{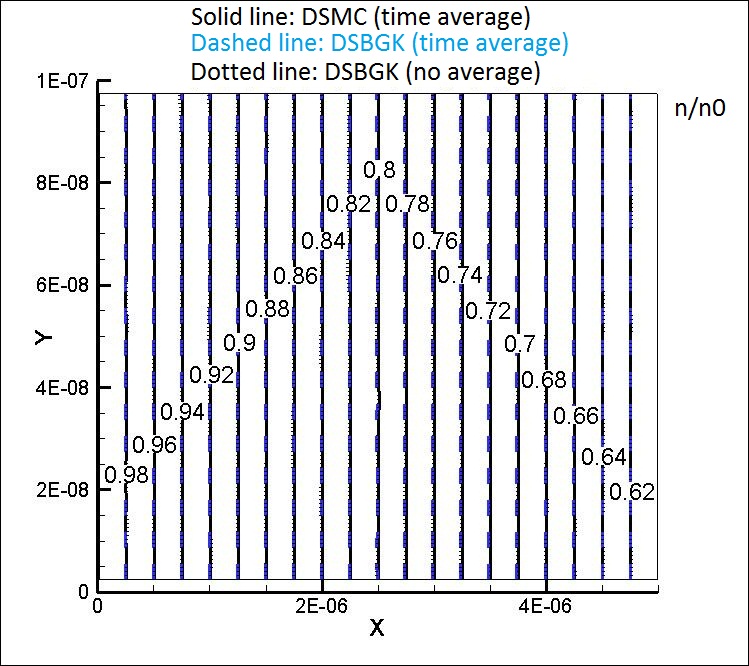}
  \includegraphics[width=0.45\textwidth]{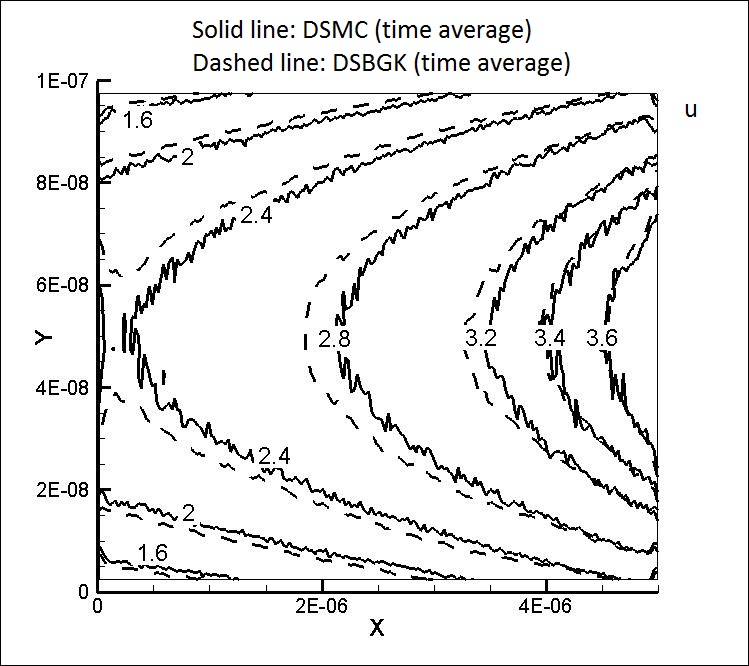} \\
  \includegraphics[width=0.45\textwidth]{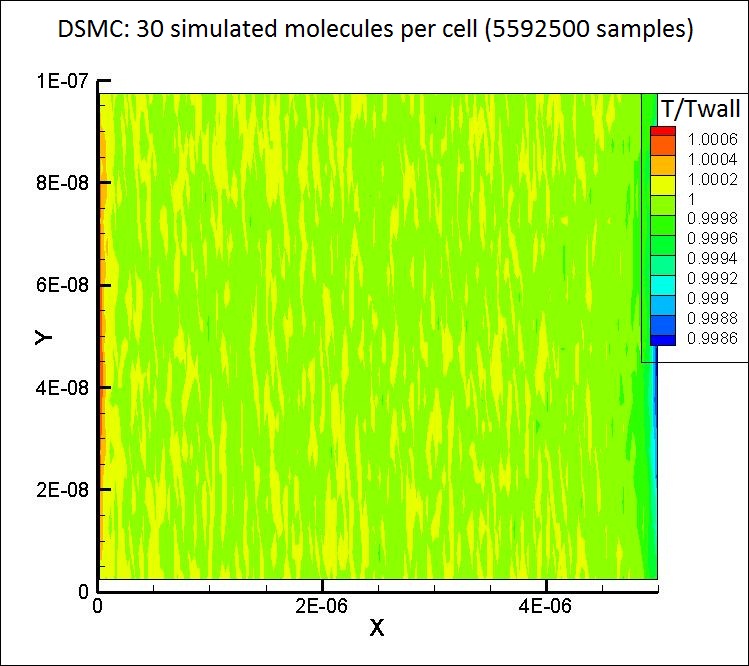}
  \includegraphics[width=0.45\textwidth]{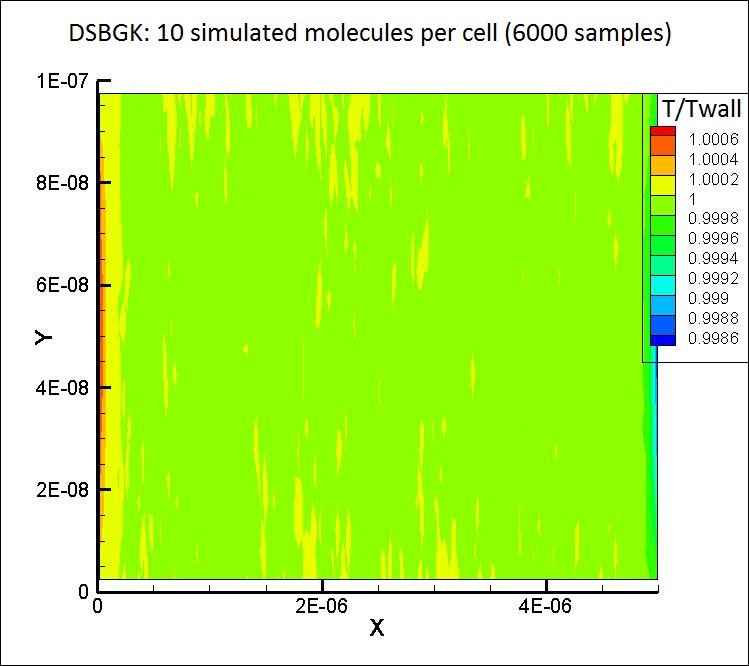} \\
  \includegraphics[width=0.45\textwidth]{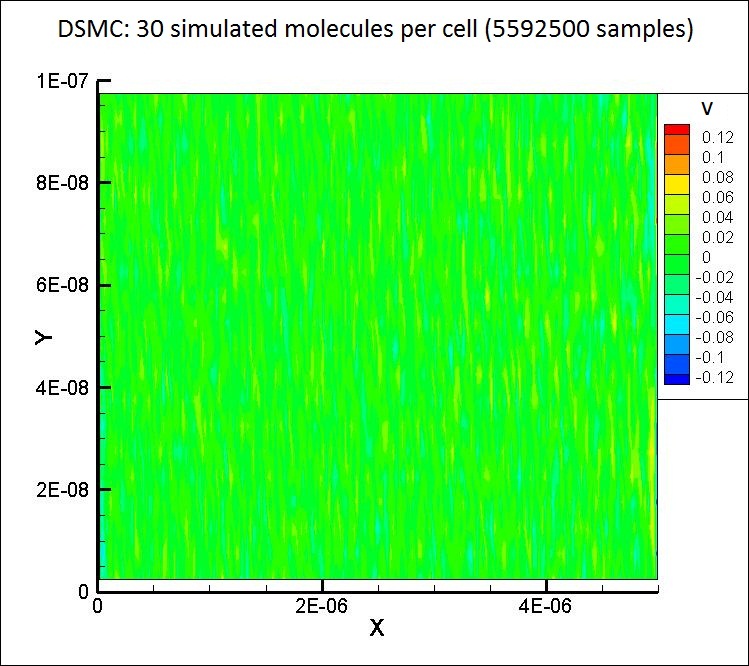}
  \includegraphics[width=0.45\textwidth]{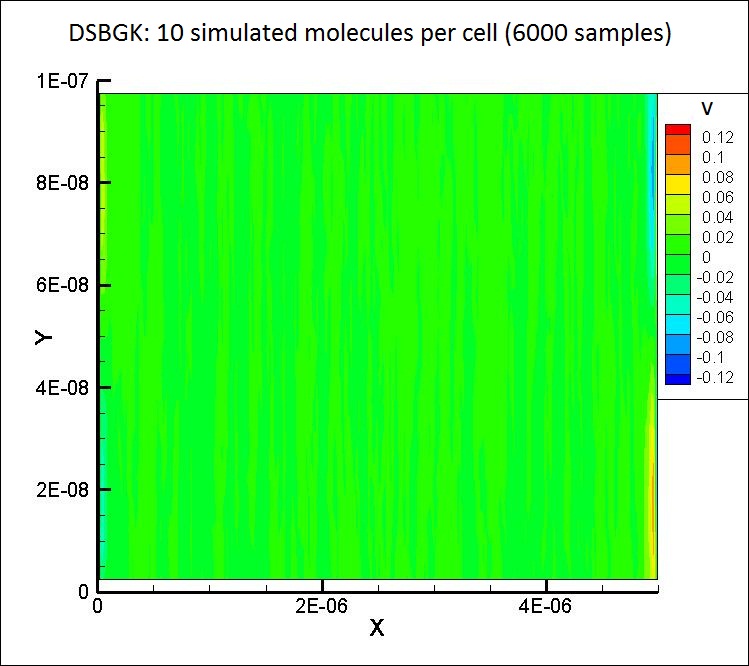} \\
  \caption{Comparison between DSMC and DSBGK methods in channel flow problem, $Kn=0.63$ and $n_\mathrm{outlet}$=0.6$n_0$.}
  \label{fig:ChannelflowKn0.63}
\end{figure}

After convergence, the transient DSBGK results at $30000^{th} \Delta t$ are given in Fig. \ref{fig:ChannelflowDSBGK} taking about 36 minutes of computational time. It shows that the DSBGK simulation is stable when using only 10 simulated molecules per cell in open problem. Using few simulated molecules reduces the memory usage and improves the applicability in problems of large scale. The transient $n$ is smooth but the transient $T, u, v$ contain large stochastic noise. We use the time-average process to reduce noise and collect 6000 samples (sampling at each $\Delta t$) after $30000 \Delta t$, which takes about 8 minutes making the total computational time about 44 minutes. The time-average results of DSBGK simulation are given in Fig. \ref{fig:ChannelflowKn0.63} with comparison by the time-average results of DSMC simulation. Unfortunately, the average results of $v$ and $T$ of the DSBGK and DSMC simulations are still dominated by the stochastic noise due to small variations inside the flow domain, particularly in the area far away from the two ends. It should be pointed out that the average $v$ and $T$ can distinctly show their main variations near the inlet and outlet. Note that the dominance of stochastic noise is due to not only small characteristic velocity but also small variation. As we can see from the lid-driven problem at small driven velocity $U_\mathrm{wall}=0.1$ m/s, the transient velocity distribution is smooth during the whole evolution process as its variation inside the whole flow domain is obvious (see Fig. \ref{fig:CavityflowDSBGK}). The magnitude of stochastic noise in the DSBGK time-average results using a small sample size is much smaller than that in the DSMC time-average results using a large sample size. In addition, the agreement between the DSBGK time-average and DSBGK transient results of $n$ implies that the nonphysical fluctuation of $n$ observed in the DSBGK simulation of closed problem is eliminated in the open problem and the time-average process is valid for sampling $n$ if necessary.

\section{Conclusions}\label{s:conc}
The DSMC algorithm is analyzed using the importance sampling scheme to solve the Boltzmann equation. The DSBGK algorithm is introduced by theoretical analysis which shows the convergence of DSBGK method to the BGK equation. Many numerical results in several benchmark problems are listed together to show the validity and high efficiency of the DSBGK method.

Unsolved Problems in the current DSBGK algorithm include: 1) hysteresis effect in transient problems; 2) nonphysical fluctuation of density distribution in \textit{closed} problems; 3) how to get a rigorous formula of $f_\mathrm{B,CL}(\vec c_\mathrm{r})$ using Eqs. \eqref{eq:kernel} and \eqref{eq:kernelCLL}.

\end{document}